\documentclass[preprint]{emulateapj}

\usepackage{amsmath}
\usepackage{amssymb}
\usepackage{bm}
\usepackage{ulem}
\usepackage[table]{xcolor}
\colorlet{tablegray}{gray!25}

\slugcomment{draft v4}

\shorttitle{}
\shortauthors{Odaka et al.}

\begin{document}


\title{Short-Term Variability of X-rays from Accreting Neutron Star Vela X-1: I. Suzaku Observations}


\author{Hirokazu Odaka\altaffilmark{1}, Dmitry Khangulyan\altaffilmark{1}, Yasuyuki T.\ Tanaka\altaffilmark{2}, Shin Watanabe\altaffilmark{1}, \\  Tadayuki Takahashi\altaffilmark{1,3}, and Kazuo Makishima\altaffilmark{3}}


\altaffiltext{1}{Institute of Space and Astronautical Science (ISAS), Japan Aerospace Exploration Agency (JAXA), 3-1-1 Yoshinodai, Chuo, Sagamihara, Kanagawa, 252-5210, Japan}
\altaffiltext{2}{Hiroshima Astrophysical Science Center, Hiroshima University, 1-3-1 Kagamiyama, Higashi-Hiroshima, Hiroshima 739-8526, Japan}
\altaffiltext{3}{Department of Physics, University of Tokyo, 7-3-1 Hongo, Bunkyo, Tokyo, 113-0033, Japan}

\newcommand{\degree}{$^\circ$~}


\begin{abstract}
We have analyzed the time variability of the wide-band X-ray spectrum of Vela X-1, the brightest wind-fed accreting neutron star, on a short timescale of 2 ks by using {\it Suzaku} observations with an exposure of 100 ks.
During the observation, the object showed strong variability including several flares and so-called ``low states'', in which the X-ray luminosity decreases by an order of magnitude.
Although the spectral hardness increases with the X-ray luminosity, the majority of the recorded flares do not show any significant changes of circumstellar absorption. However, a sign of heavy absorption was registered immediately before one short flare that showed a significant spectral hardening.
In the low states, the flux level is modulated with the pulsar spin period, indicating that even at this state the accretion flow reaches the close proximity of the neutron star. Phenomenologically, the broad-band X-ray spectra, which are integrated over the entire spin phase, are well represented by the ``NPEX'' function (a combination of negative and positive power laws with an exponential cutoff by a common folding energy) with a cyclotron resonance scattering feature at 50 keV.
Fitting of the data allowed us to infer a correlation between the photon index and X-ray luminosity.
Finally, the circumstellar absorption shows a gradual increase in the orbital phase interval 0.25--0.3, which can be interpreted as an impact of a bow shock imposed by the motion of the compact object in the supersonic stellar wind.

\end{abstract}




\bibliographystyle{apj}


\section{Introduction}


Accretion-powered pulsars in high-mass X-ray binaries (HMXBs) are the best laboratories for studying the accretion in the presence of a strong magnetic field.
Such binary systems harbor neutron stars with a typical magnetic field of $\sim$$10^{12}\ \mathrm{G}$, which accrete matter from the surrounding environment, and their gravitational energy is realized through the bright X-ray radiation.
In spite of intensive studies of the X-ray pulsars since the dawn of X-ray astronomy, their emission mechanism still remains unclear due to the extreme physical conditions induced by strong gravitation and magnetic fields.
Given the shape of the X-ray spectrum, which can be described by a hard power law with a quasi-exponential cutoff at an energy of 20-30 keV, Comptonization by hot electrons in the accretion flow can provide a feasible scenario for the formation of the X-ray spectral continuum \citep{Sunyaev:1980, Makishima:1999}.
However, the relation between the radiation and the physical conditions of the accretion flow has not been understood yet.


In order to confirm the hypothesis of the Comptonized emission, it is essential to build more detailed physical models of the accreted plasma that radiates X-rays.
Since the spectral variability reflects changes of the physical conditions of the plasma, the variability patterns can contain important information about the accreted plasma.
We therefore analyzed X-ray archive data of HMXB Vela X-1 obtained with the {\it Suzaku} X-ray observatory to pursue the variability of hard X-rays from the central engine.
The first analysis results of this data are presented by \citet{Doroshenko:2011}.
Owing to the wide-band coverage and the low-background capability of {\it Suzaku} \citep{Mitsuda:2007}, we are able to obtain unprecedentedly high-quality spectra with a short integration time of 2~ks.
Furthermore, the Hard X-ray Detector (HXD) \citep{Takahashi:2007, Kokubun:2007} onboard {\it Suzaku} is the only instrument that is able to obtain such a short-period spectrum in the hard X-ray band above 10 keV with high signal-to-noise ratios.


Vela X-1 (4U 0900$-$40) is the best target for such study since it is the brightest wind-fed pulsar, which always displays strong time variability of the X-ray flux level.
In particular, the source displays intensive outbursts, e.g., the flaring behavior in hard X-rays was observed with {\it INTEGRAL} up to an intensity of $\sim 5\ \mathrm{Crab}$ at 20--40 keV
\citep{Kreykenbohm:2008}.
The binary consists of a B0.5Ib supergiant HD 77581 of radius $30.0R_\odot$, and a neutron star with a spin period of 283.5 seconds \citep{McClintock:1976, Kreykenbohm:2008} on a nearly circular orbit with a small binary separation of $53.4R_\odot$ \citep{vanKerkwijk:1995} and orbital period of 8.964 days \citep{vanKerkwijk:1995}.
Since the B-type companion drives a strong stellar wind with a mass-loss rate of about $2\times 10^{-6} M_\odot \mathrm{yr}^{-1}$ \citep{Watanabe:2006} and terminal velocity of 1100 $\mathrm{km\ s^{-1}}$ \citep{Prinja:1990}, the neutron star is always embedded deeply in the dense stellar wind.
This creates perfect conditions for an intensive accretion, which leads to the formation of the bright X-ray emission.
Assuming a distance of 1.9 kpc to the object \citep{Sadakane:1985}, the typical X-ray luminosity of Vela X-1 is $\sim 4\times 10^{36}\ \mathrm{erg\ s^{-1}}$.


Though it is beyond doubt that the accretion rate to the neutron star is highly variable in such systems, the mechanism that governs accretion is still in debate.
Consequently, the physical origin of the X-ray variability has not been studied in detail yet.
One straightforward, possible explanation for the time variability is that inhomogeneous structure of the stellar wind leads to variations of the density and velocity of matter around the neutron star, resulting in a fluctuation of the accretion rate \citep{Kreykenbohm:2008, Ducci:2009, Furst:2010}.
Another interpretation is that a magnetospheric barrier regulates the accretion \citep{Grebenev:2007,Bozzo:2008}.
Within this scenario, the origin of the variability was discussed in \citet{Doroshenko:2011}, who has analyzed ``off states'' of Vela X-1 based on the {\it Suzaku} data (these data are also used in this paper).


The present paper is organized as follows.
Section~\ref{sec:observation} describes the {\it Suzaku} observation of Vela X-1 and the data reduction.
The analysis results, which include the time variability of the spectrum, are presented in
Sections~\ref{sec:analysis_xis} and \ref{sec:analysis_xishxd}.
Section \ref{sec:discussion} discusses the time variability, the structure of the wind, and spectral features of the neutron star radiation.
In Section~\ref{sec:conclusions} we summarize our findings and conclusions.
Implications of the obtained observational results to detailed modeling of the central engine are presented separately by \citet{Odaka:2013b}.

\section{Observations and Data Reduction}
\label{sec:observation}

The observation of Vela X-1 with {\it Suzaku} was performed in June 2008 for an exposure time of 100 ks (Principal Investigator: T.~Kallman).
Information on the observation is summarized in Table~\ref{table:obs_summary_velax1}.  According to the ephemeris obtained with {\it INTEGRAL} \citep{Kreykenbohm:2008}, the {\it Suzaku} observation covered the orbital phase interval between 0.175 and 0.310, which corresponds to 13.5\% of the orbital period of the Vela X-1 binary system of 8.964 days.
Since the target is very bright, all of the three CCDs (XIS 0, 1, 3) were operated under the 1/4 window option, in which 256$\times$1024 pixels of the CCD are read out every 2 s.

We used the {\it Suzaku} data of Vela X-1 processed by a set of software for the {\it Suzaku} processing pipeline version 2.2.8.20, and used the {\it HEASoft} software package version 6.9 together with a calibration database updated on 2010-08-12 for data processing as well as basic analysis including spectral fitting.
For all the data with the XIS and the HXD, we reprocessed unfiltered event data by ourselves to apply new calibration information on the detector gain and to avoid a problem of the {\it cleansis} tool, which usually eliminates flickering events of the CCDs but sometimes fails with old versions due to its detection algorithm.
In the data reprocessing, standard event screening criteria were applied for both the XIS and the HXD.
These standard screening procedures include selection by event grades and removal of time intervals such as the spacecraft's passage of the South Atlantic Anomaly (SAA), regions of small geomagnetic cut-off rigidity (COR), and those of a low elevation angle from the Earth.
For the XIS, events for an elevation angle larger than 5\degree above the Earth limb and
larger than 20\degree above the sunlit Earth limb are selected.
The HXD data selection requires an elevation angle larger than 5\degree above the Earth limb and the COR larger than 6 GV.

In the following sections the observational results obtained with {\it Suzaku} from Vela X-1 and details of the data analysis performed are described.
Due to a noticeably better performance of {\it Suzaku} below 10~keV, the analysis was 
conducted in two ways: (1) soft-X-ray analysis below 10 keV using only the XIS data (\S\ref{sec:analysis_xis}); and (2) the XIS/HXD simultaneous wide-band analysis (\S\ref{sec:analysis_xishxd}).
The dedicated soft X-ray analysis, independent of any impact of the hard X-ray instrument, is also justified from a physical point of view, since the soft-X-ray properties of the Vela X-1 system are particularly useful for studying the relation between the circumstellar medium and the activity of the compact object from the viewpoint of the origin of the time variability.
On the other hand, the hard X-ray information is the key to grasp the radiation mechanism from accretion-powered pulsars.
Indeed, the spectral features of the emission from accreted plasma appear above 10 keV in the hard X-ray band, while the soft X-ray continuum has a featureless power-law
shape.
Generally, the X-ray pulsar spectrum has a quasi-exponential cutoff at 20--30 keV, and often shows absorption-like features due to the cyclotron resonance in the hard X-ray band.
The spectral analysis was performed by using a spectral fitting software package {\it Xspec} version 12.6 included in {\it HEASoft}.

\begin{table}[htdp]
\caption{Summary of the {\it Suzaku} observation of Vela X-1}
\begin{center}
\begin{tabular}{ll}
\hline\hline
Property & Value \\
\hline
Obs.~ID & 403045010 \\
Coordinates (RA, Dec) (J2000) & ($135.5377$, $-40.5514$) \\
Pointing & XIS nominal \\
Start time & 2008-06-17 04:45:17 \\
Start time (MJD) & 54634.19811 \\
Observation duration & 145 ks \\
Orbital phase & 0.175 - 0.310 \\
Exposure time & 104 ks \\
\hline
XIS clocking mode & Normal \\
XIS burst option & No \\
XIS window option & 1/4 window \\
XIS Spaced-row Charge Injection & On \\
\hline
\end{tabular}
\end{center}
\label{table:obs_summary_velax1}
\end{table}%


\section{Soft X-Ray Data Analysis}
\label{sec:analysis_xis}

In this section we present analysis of data obtained in the soft X-ray energy band with XIS of the {\it Suzaku} telescope.
The performed analysis includes a study of the pileup effect in the data (\S~\ref{subsec:xis_pre_analysis}), of the temporary behavior of the source (\S~\ref{subsec:xis_lc}), and of spectral properties (\S~\ref{subsec:xis_spectrum}).

\subsection{Preparatory Analysis and Photon Pileup Effects}
\label{subsec:xis_pre_analysis}

Primarily, we have examined the effects of photon pileup on the CCDs.
The photon pileup effect is important in the cases if the observed source is very bright: two or more photons can strike the same CCD pixel during a single readout and generate an event
regarded as a non-X-ray event or a higher energy photon.
This results in an inaccurate estimate of the spectral properties and source flux.
The modest angular resolution of the {\it Suzaku} X-ray mirrors and a short exposure of 2 seconds in a single readout of the CCD with the 1/4 window option do reduce the pileup effect.
However, since Vela X-1 is an extremely bright source, a careful treatment of the possible photon pileup is necessary.

\begin{figure}[tbp]
\begin{center}
\includegraphics[width=9cm]{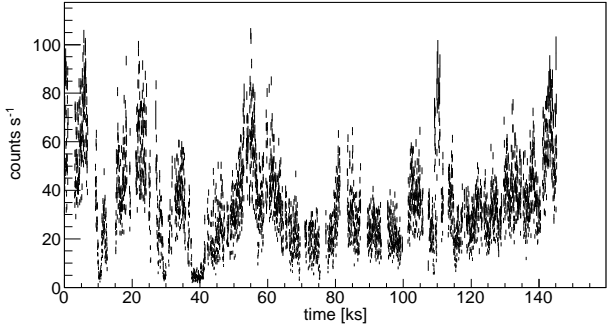}
\caption{Light curve of Vela X-1 obtained with XIS 0 (0.3-12 keV). The time bin size is 24 s.}
\label{fig:xis0_lc}
\end{center}
\end{figure}

First, we have extracted a light curve from the circular region centered on the image peak with a radius of $120''$ to check the significance of the pileup effect and temporal behavior of the source.
The light curve, shown in Figure~\ref{fig:xis0_lc}, displays strong time variability on top of the X-ray pulsation with a period of 284 s.
The count rate reaches about $100\ \mathrm{counts\ s^{-1}}$ in periods of the flaring behavior.
Since operations of the XIS with the ``1/4 window'' and ``no burst'' options requires a count rate less than $50\ \mathrm{counts\ s^{-1}}$ to avoid a strong pileup, the photon events recorded from the Vela X-1 can be considerably affected by the pileup effects.

A simple standard method to correct the pileup effect is to remove the center part from the event extraction region for light curves or spectra.
In order to find an appropriate extraction region, we performed preparatory spectral analysis of the XIS data. We defined five extraction regions in a shape of a circle or annulus: (1) $0<r<120''$ (no removal of the image center); (2) $20''<r<120''$; (3) $30''<r<120''$; (4) $45''<r<120''$; and (5) $60''<r<120''$.
Here $r$ denotes the radial distance from the image center.
From these regions, we extracted spectra of the XIS every 2268 s, which corresponds to eight times the spin period of the X-ray pulsar.
Background spectra were extracted from source-free regions.  Detector response matrices (``rmf'') of the XIS were generated by {\it xisrmfgen}, and an ancillary response function (``arf'') was simulated by {\it xissimarfgen} for each extraction region and each time interval.

It is well-known that soft X-ray emissions from Vela X-1 below 10 keV can be characterized by an absorbed power law with a strong iron fluorescence line at 6.4 keV.
Thus, we modeled the XIS spectra by fitting to
\begin{equation}\label{eq:model_xis_pre}
\frac{dN}{dE} = \exp(-N_\text{H}\sigma_\text{abs}(E))(AE^{-\Gamma}+F_\text{Fe}(E)),
\end{equation}
where $E$, $N_\text{H}$, $\sigma_\text{abs}(E)$, $A$, $\Gamma$ are photon energy, equivalent hydrogen column density of absorbing matter, cross section of the photoelectric absorption on hydrogen, normalization, and power law exponent, respectively.
Here, we used {\it wabs} model in {\it Xspec} for the absorption model.
The iron line was modeled by a Gaussian $F_\text{Fe}(E)$ with a very narrow width at 6.4 keV.  In the fitting, $N_\text{H}$, $\Gamma$, $A$ and the Gaussian normalization were treated as free parameters.
It is also known that the emission scattered from an ambient accretion flow or stellar wind sometimes results in a strong soft X-ray excess (below $\sim$2~keV) on top of the absorbed power-law continuum.
To reduce effects of the soft excess on spectral fitting, we restricted the energy range for fitting to 2.5--10 keV.

\begin{figure*}[tbp]
\begin{center}
\includegraphics[width=16.5cm]{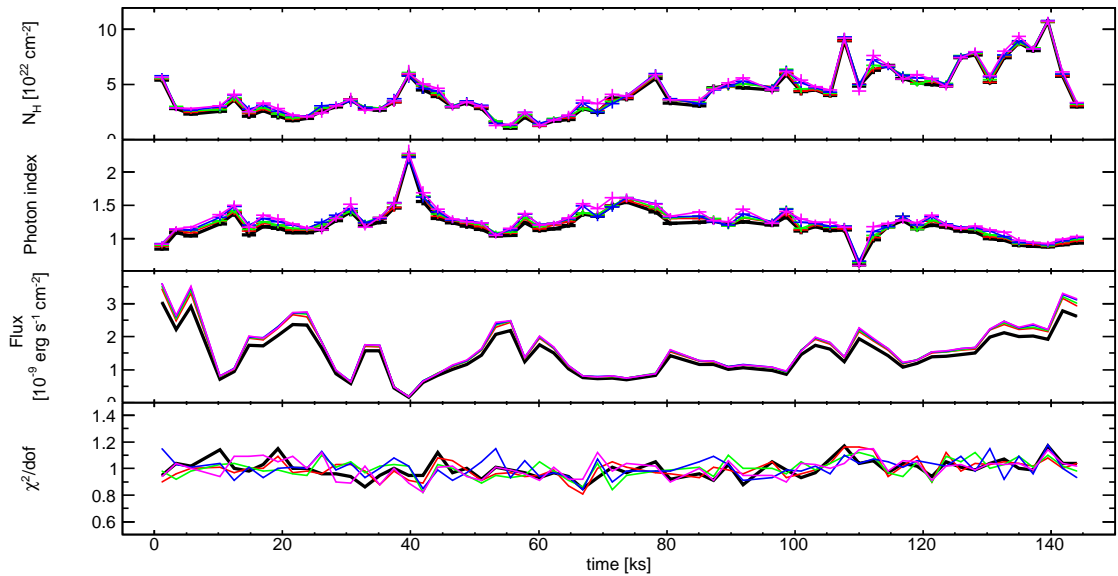}
\caption[Best-fit parameters of the XIS 0 spectra as functions of time
for different extracting regions]{Best-fit values of the equivalent
  hydrogen column density $N_\text{H}$ and the photon index $\Gamma$
  obtained from the XIS 0 spectra as functions of time for different
  extracting regions together with calculated flux and reduced
  $\chi^2$ of the fit. The parameters from the extraction regions of
  $0<r<120''$ (black), $20''<r<120''$(red), $30''<r<120''$(green),
  $45''<r<120''$(blue), and $60''<r<120''$(magenta) are shown. The
  error bars with $N_\text{H}$ and photon index $\Gamma$ show
  1$\sigma$ uncertainties.}
\label{fig:xis0_fit_pl}
\end{center}
\end{figure*}

The fitting results for XIS 0 are shown in Figure~\ref{fig:xis0_fit_pl} for the whole observation period.  The spectra extracted from the region that includes the image center (black line) show lower fluxes and a harder spectrum particularly at epochs of high flux level, as compared to the spectra extracted from other regions.
This is a straightforward indication of photon pileup.
The effects become less significant when the central region excluded from the data analysis is increased.
We considered that the region of $45''<r<120''$ (blue lines in the figure) is suitable for our spectral and timing analysis given  that the pileup effects are negligible and this region contains enough events to be statistically significant.
The analysis of the XIS data described in the following sections is based on the events extracted from the region of $45''<r<120''$ for all the CCDs.

\subsection{Light Curves}\label{subsec:xis_lc}

\begin{figure}[htbp]
\begin{center}
\includegraphics[width=9cm]{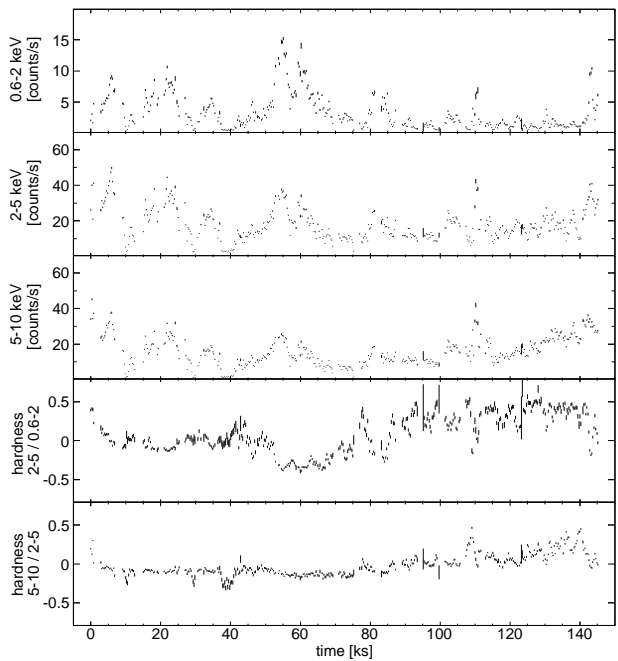}
\caption{Light curves from XIS for three different energy ranges of
  0.6--2 keV, 2--5 keV, and 5--10 keV. Hardness ratios between the
  different energy ranges are also presented. The data count rates
  were obtained by summing those of XIS 0 and XIS 3. The time bin size
  is 284 s, the same as the spin period of the neutron star.}
\label{fig:xis_lc_284s}
\end{center}
\end{figure}

Since the X-ray emission from Vela X-1 displays a strong time variability in addition to the X-ray modulation caused by the pulsar (i.e., that due to the pulsar rotation), it is essential for understanding of the emission nature to examine the obtained light curves.
In Figure~\ref{fig:xis_lc_284s}, we show light curves obtained with the XIS for different energy ranges: 0.6--2 keV, 2--5 keV, and 5--10 keV.
The light curves were extracted by summing up the count rates of XIS 0 and XIS 3.
The light curve time bin was selected to coincide with the spin period of the neutron star, i.e. 284~s.  Since the background levels are negligible and no flaring of the background occurred during the observation, no background subtraction is necessary.

The count rates in the three different energy ranges provide us with the spectral evolution of Vela X-1.  The lowest-energy band (0.6--2 keV) is heavily affected by photoelectric absorption in the accretion flow or stellar wind, while the higher energy bands are not significantly affected by the absorption.
In order to characterize the spectral properties at each moment, we have introduced the hardness ratio between the hard and soft bands as
\begin{equation}
\text{hardness ratio} = \frac{h-s}{h+s}=\frac{H/\bar{H}-S/\bar{S}}{H/\bar{H}+S/\bar{S}}\,.
\end{equation}
Here, $h$ and $s$ are count rates normalized to the average count rates in the hard and soft bands, respectively; $H$ and $S$ are the corresponding rates in these bands, and the bar indicates the value averaged over the whole observation.
This hardness ratio can be used as an indicator of the spectral hardness, ranging between $-1$ and $+1$.
In addition to the light curves of count rate, we show the hardness ratios for two different sets of energies in Figure~\ref{fig:xis_lc_284s}.
The hardness ratio calculated for the bands of 0.6--2 keV and 2--5 keV is sensitive to the amount of absorbing medium, or $N_\text{H}$.
On the contrary, the ratio for 2--5 keV and 5--10 keV is sensitive to the intrinsic spectral slope rather than to the absorption impact.

\begin{figure}[htbp]
\begin{center}
\includegraphics[width=9cm]{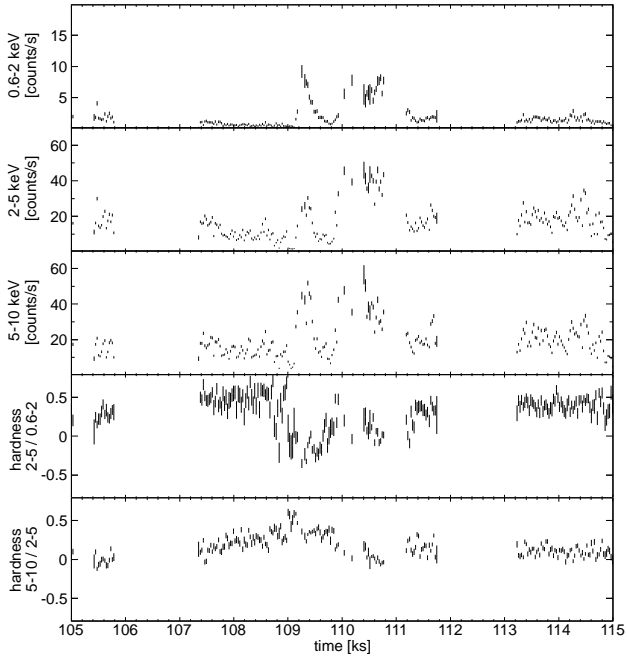}
\caption{Light curves and hardness ratios from XIS around the flare
  at $t=110$ ks. The time bin size is 28 s. Other conditions of the
  light curve extraction are the same as
  Figure~\ref{fig:xis_lc_284s}.
  At the two long data blanks from $t=106.8$ ks and $t=111.8$ ks, the satellite passed in the SAA, with the instruments turned off. Absence of data points that is seen between $t=107.4$ ks and $t=111.8$ ks is due to data removal by saturation of the satellite telemetry or by a condition of an elevation angle smaller than 5\degree above the Earth limb.}
\label{fig:xis_lc_flare110ks}
\end{center}
\end{figure}

During the 140-ks-long observation of Vela X-1, several flares, which were a few times brighter than the average flux level, have been detected.
Most of them had a duration of $\sim$10 ks and the hardness ratio does not show any significant spectral changes during these flares.
In contrast, the flare at $t=110$ ks had different characteristics: it has been characterized by a very short (about 1 ks) duration accompanied by a significant spectral hardening.
Figure~\ref{fig:xis_lc_flare110ks} shows light curves around the flare with short time bins of 28 s.
It can be seen that the flare around $t=110$~ks consisted of two shorter sub-flares at $t=109.3$ ks and
$t=110.4$ ks.
The former displayed a significant spectral hardening while the hardness ratio of the latter was similar to the typical or average value ($\sim$0.0).
Interestingly, a strong attenuation of the soft band, probably caused by heavy absorption, was observed just before the first sub-flare.

\begin{figure}[htbp]
\begin{center}
\includegraphics[width=9cm]{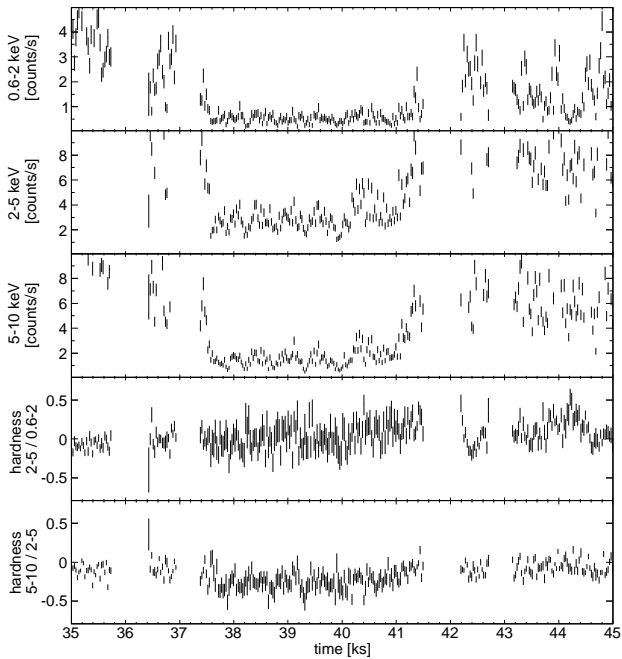}
\caption{Light curves and hardness ratios of the XIS around the low
  state at $t=40$ ks. Vela X-1 dropped into the low state from $t=38$
  ks until $t=41$ ks. The time bin size is 28 s. Other conditions of
  the light curve extraction are the same as
  Figure~\ref{fig:xis_lc_284s}.}
\label{fig:xis_lc_offstate40ks}
\end{center}
\end{figure}

Besides the flaring behavior, the object sometimes dropped into ``low states'', in which the count rates were about an order of magnitude smaller than the average.
Three periods of the ``low states'' are seen in Figure~\ref{fig:xis_lc_284s}: $t=$10 ks, 30 ks, and 40 ks.  The ``low state'' observed at $t=40$ ks lasted for 3.5 ks, while the duration of others was about 1 ks.
Apparently, a spectral softening was occurring during all the ``low states''.  This spectral feature might be essential to understand the physical situation of X-ray pulsars in ``low states''.
Figure~\ref{fig:xis_lc_offstate40ks} shows light curves around the low state at $t=40$ ks with short time bins of 28 s.
In the low state, Vela X-1 still displays X-ray pulsations, clearly indicating that the accretion flow reaches near the magnetic poles on the stellar surface.
\citet{Doroshenko:2011} reported detailed data analysis and its interpretation based on the same {\it Suzaku} data of these low states.

After $t=115$ ks (until the end of the observations), the count rates in the hard bands were gradually increasing, while the soft band (0.6--2 keV) did not show such a growth.
This likely implies that the soft emission has been heavily absorbed by dense matter located between the compact object and the observer.
Large values of the spectral hardness between the two softer bands during that period also suggest strong absorption.
Fast changes of this hardness ratio seen during that epoch can be caused by fluctuations of the absorbing matter density.

\subsection{Spectral Analysis}\label{subsec:xis_spectrum}

In order to evaluate the spectral evolution of Vela X-1 quantitatively, we have performed time-resolved spectral analysis.
We extracted spectra and generated the detector response functions in the way described in \S\ref{subsec:xis_pre_analysis}.
Except for the soft excess due to the photoionized stellar wind, the X-ray emission below 10 keV is well described by an absorbed power law with iron fluorescence.
Since in the preparatory analysis (\S\ref{subsec:xis_pre_analysis}) we have found the absorption edge of iron at $E_\text{edge}\sim$7.1~keV, we model the soft X-ray emission by
\begin{equation}\label{eq:spec_model_xis}
\frac{dN}{dE} = \exp(-N_\text{H}\sigma_\text{abs}(E))f_\text{edge}(E)(AE^{-\Gamma}+F_\text{Fe}(E)).
\end{equation}
The additional factor $f_\text{edge}(E)$ to the preliminary model described by Equation (\ref{eq:model_xis_pre}) represents the effect of additional absorption by iron, written as
\begin{equation}
f_\text{edge}(E) = \left\{ \begin{array}{ll}
1 & (E < E_\text{edge}) \\
\exp[-\tau_\text{edge}(E/E_\text{edge})^{-3}] & (E \ge E_\text{edge}),
\end{array} \right.
\end{equation}
where $\tau_\text{edge}$ denotes the optical depth for the iron absorption.

We performed spectral fitting of the XIS spectra for intervals of 2268~s, which correspond to eight spin periods of the neutron star.
Each spectrum has been extracted from the region of $45''<r<120''$ to avoid the photon pileup effect.
In the fitting, we have assumed  $N_\mathrm{H}$, $\tau_\text{edge}$, the power-law index $\Gamma$, power-law normalization $A$, and the Fe line normalization to be free parameters.
The energies of the iron line center and the iron K-edge were fixed to 6.4 keV and 7.25 keV, respectively, as have been determined by preliminary fitting.
The energy range for fitting has been restricted to 2.5--10 keV to avoid the impact of the soft excess in the low energy band, as we mentioned in \S\ref{subsec:xis_pre_analysis}.

\begin{figure*}[tbp]
\begin{center}
\includegraphics[width=16.5cm]{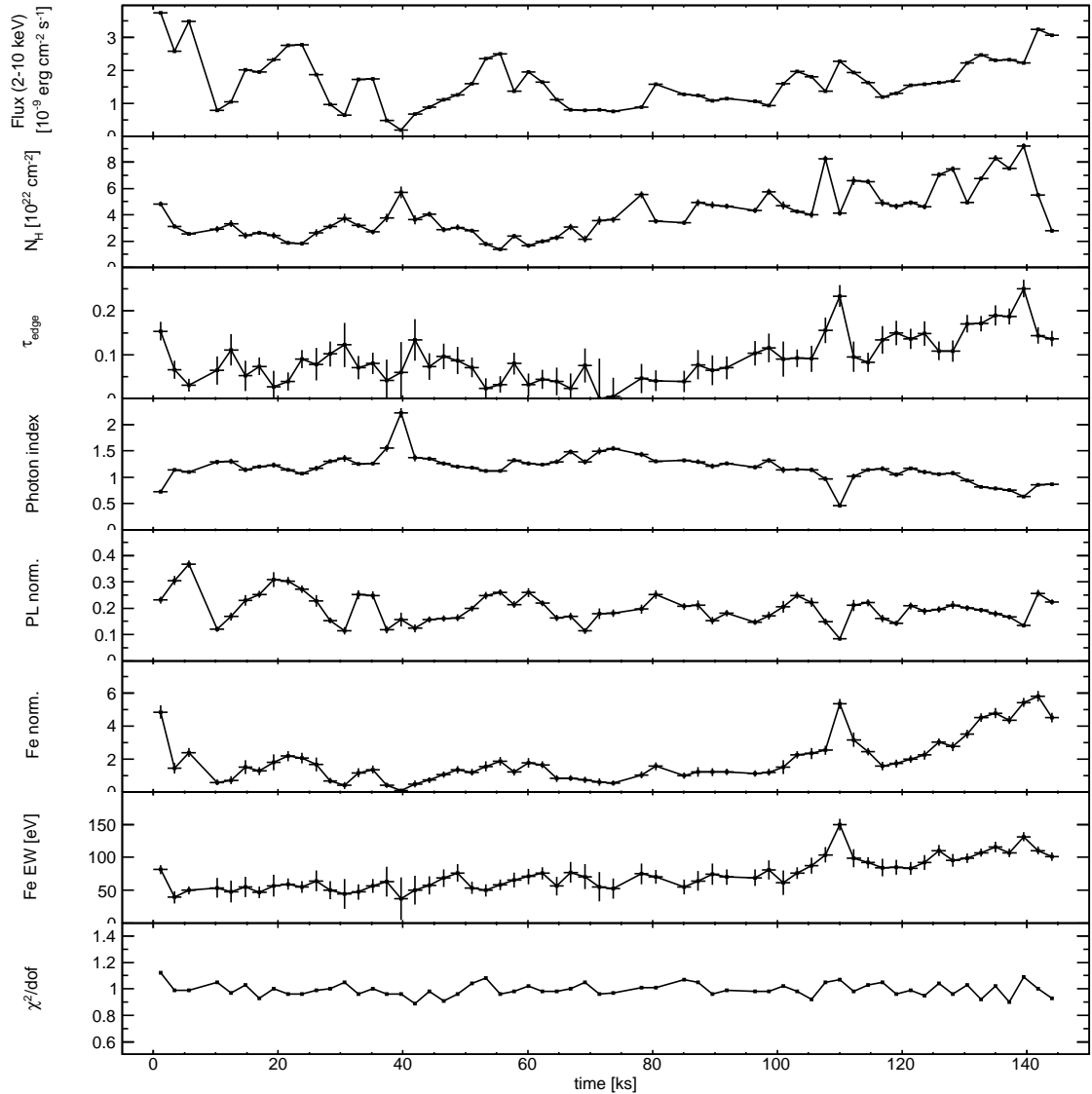}
\caption{Best-fit parameters of the XIS data to power-law models as
  functions of time. The values are tabulated in
  Table~\ref{table:fit_xis_pl}. The error bars represent 1$\sigma$
  uncertainties.}
\label{fig:plot_fit_xis_pl}
\end{center}
\end{figure*}

The results of the spectral fitting are tabulated in Table~\ref{table:fit_xis_pl}, and their time evolutions are plotted in Figure~\ref{fig:plot_fit_xis_pl}.
Reduced $\chi^2$ values for each fit are also shown to check goodness of the fit.
We excluded a few spectra with exposure time shorter than 568 s from the fitting analysis to avoid large uncertainties due to low statistics and potential bias by spin-phase dependent effects.
This quantitative analysis confirms the behaviors of the X-ray radiation from Vela X-1 seen in the light curves, as discussed previously.

The X-ray flux displayed strong variability while the power law  index, $\Gamma$, remained relatively stable, with its value close to $\Gamma\sim1$.
Typically during the observation, the flux ranged between $1\times 10^{-9}$ and $3\times 10^{-9} \mathrm{erg\ s^{-1}cm^{-2}}$, corresponding to $\sim 100$ mCrab in this band.
As seen in the light curve (Fig.~\ref{fig:xis_lc_284s}), the object dropped into a very low-luminosity state with a flux of $1.8\times 10^{-10} \mathrm{erg\ s^{-1}cm^{-2}}$ at $t=40$ ks, showing a significant spectral softening to $\Gamma=2.2$.
In contrast, a sudden hardening to $\Gamma=0.5$ occurred at $t=110$ ks. Relatively hard states with $\Gamma\sim 0.7$ also appeared at $t=0$ ks and around $t=140$ ks, when the flux was high.

The equivalent hydrogen column density $N_\mathrm{H}$ estimated by X-ray attenuation is a good probe of the circumstellar matter in the Vela X-1 system.
During the observation, which corresponds to an orbital phase range of 0.175--0.310, the obtained $N_\mathrm{H}$ value was showing strong variability between $1\times 10^{22}$ and $9\times 10^{22}\ \mathrm{cm^{-2}}$.
Since the interstellar absorption to Vela X-1 can be estimated as $N_\mathrm{H}\sim 6\times 10^{21}\
\mathrm{cm^{-2}}$ (assuming the distance of $D=1.9$ kpc and a typical hydrogen density of $1\ \mathrm{cm^{-3}}$), the time variability of $N_\mathrm{H}$ supports inhomogeneity of the absorbing
matter inside the binary system.
The strong stellar wind of the companion star, the accretion flow to the neutron star, and the accretion wake behind the neutron star can give significant contributions to the intrinsic absorption in the system.

The information about the circumstellar matter inside the system is also reflected in the iron K$\alpha$ line at 6.4 keV, since the line is formed due to the reprocessing of the hard X-ray emission by matter.
Roughly, the line intensity is proportional to the illuminating hard X-rays from the neutron star, and the line
equivalent width is suitable for extraction of the information on the illuminated environment around the compact object.
Though the equivalent widths have relatively large uncertainties, they seem to be correlated with $N_\mathrm{H}$.
The equivalent width became large, approaching a value of $\sim150$~eV, around the epoch of $t=110$ ks, when the spectrum shows the hardening, implying an increase of the matter density near the compact object.

\begin{figure}[tbp]
\begin{center}
\includegraphics[width=4.25cm]{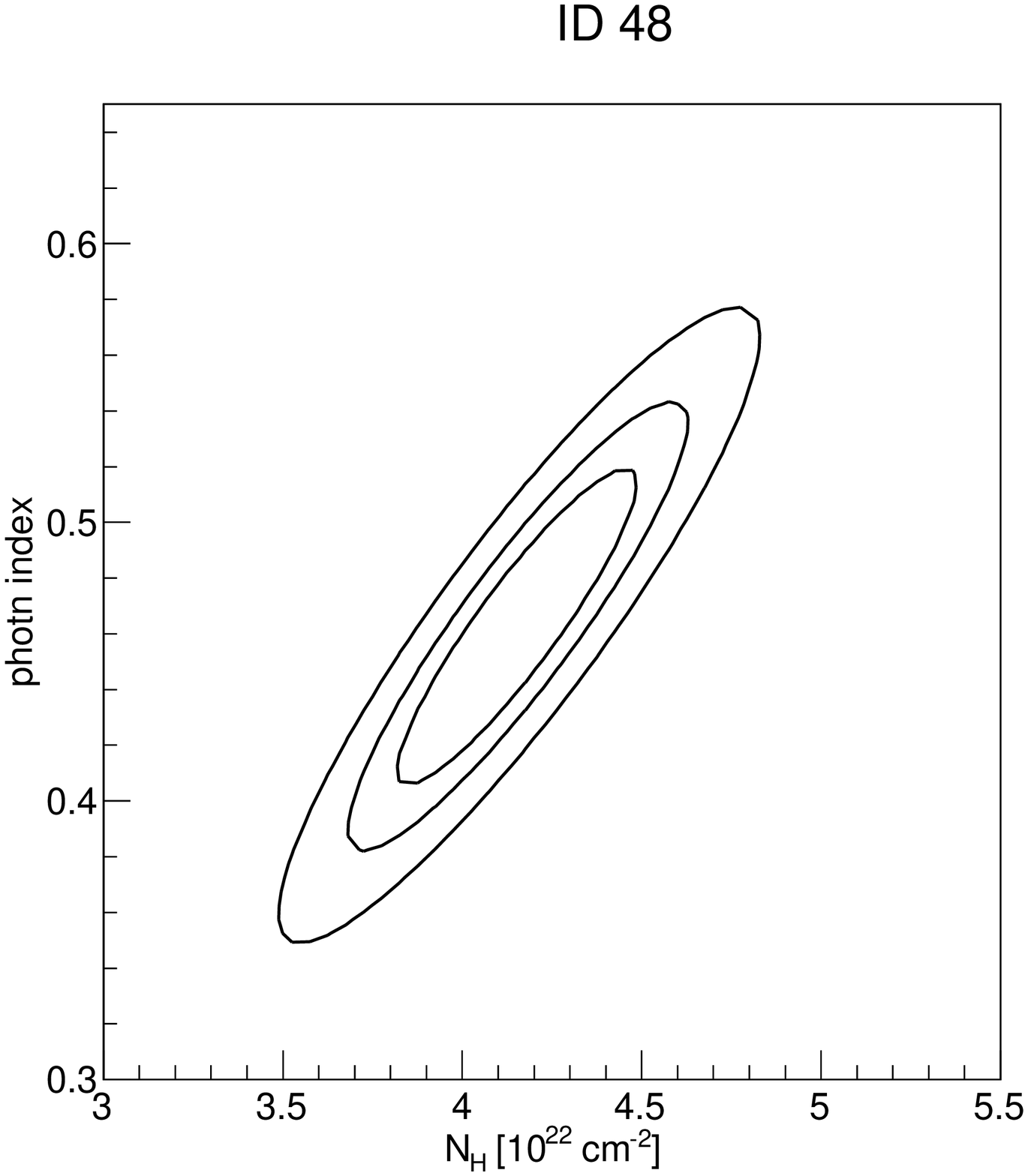}
\includegraphics[width=4.25cm]{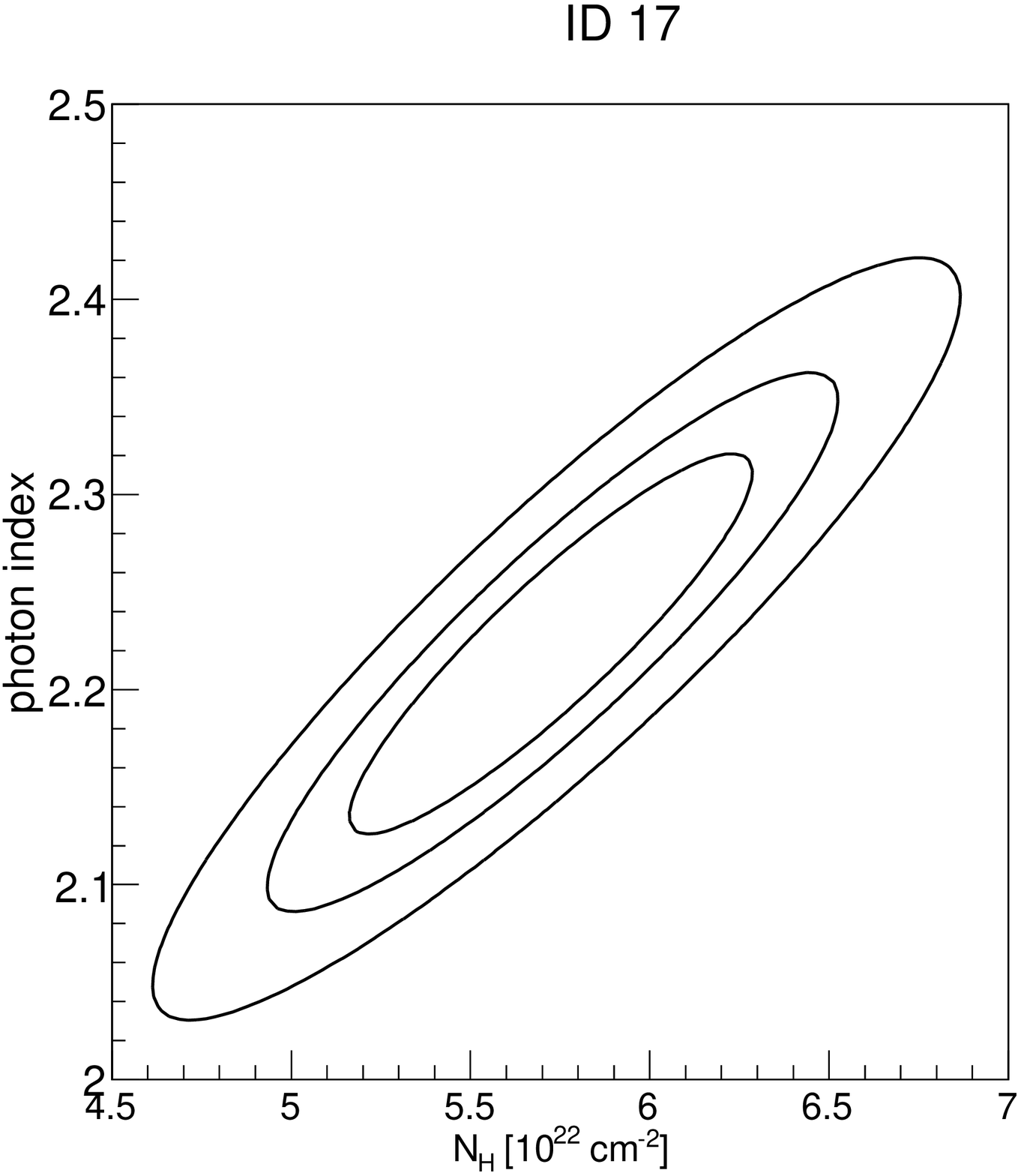}
\caption{Error contours in the $\Gamma$--$N_\mathrm{H}$ spaces for the spectra ID 48 (left) and ID 17 (right), which showed significant hardening and softening, respectively. The three contours are for 68\%, 90\%, and 99\% confidence levels.}
\label{fig:error_contour_xis_pl}
\end{center}
\end{figure}

Since the absorption hardens the spectral shape, the estimated values of the photon index $\Gamma$ and the column density $N_\mathrm{H}$ may be strongly coupled.
Therefore, the compensating behaviors of $N_\mathrm{H}$ and $\Gamma$ might be a result of wrong fitting due to the coupling.
To study this we have selected the significant hardening at $t=110$ ks (ID 48) and the significant softening at $t=40$ ks (ID 17), the two most prominent features detected for the photon index.
For these two cases, we examined the $\chi^2$ structures in the two-parameter ($\Gamma$--$N_\mathrm{H}$) space, and show the error contours for these fits in Figure~\ref{fig:error_contour_xis_pl}.
Although the two parameters have a correlation, the 99\% confidence intervals of them are significantly smaller than the temporal changes of them at the significant hardening and softening.
Thus, we conclude that the significant spectral variations are real, i.e., are not a result of the parameter coupling.

\begin{figure}[tbp]
\begin{center}
\includegraphics[width=7.0cm]{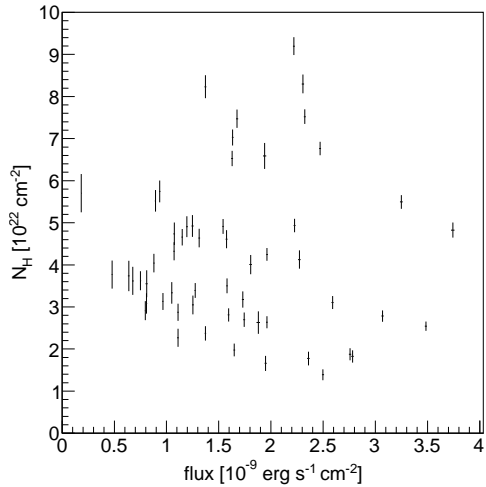}
\caption{Relation between $N_\mathrm{H}$ and the X-ray flux.}
\label{fig:plot_corr_xis_flux_nh}
\end{center}
\end{figure}

\begin{figure}[tbp]
\begin{center}
\includegraphics[width=7.0cm]{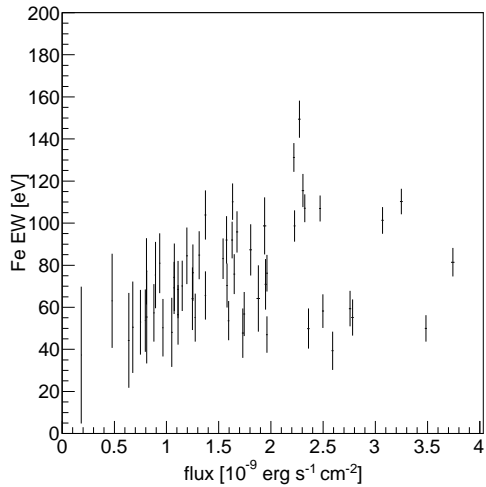}
\caption{Relation between the equivalent width of iron K$\alpha$ line
  and the X-ray flux.}
\label{fig:plot_corr_xis_flux_ew}
\end{center}
\end{figure}

\begin{figure}[tbp]
\begin{center}
\includegraphics[width=7.0cm]{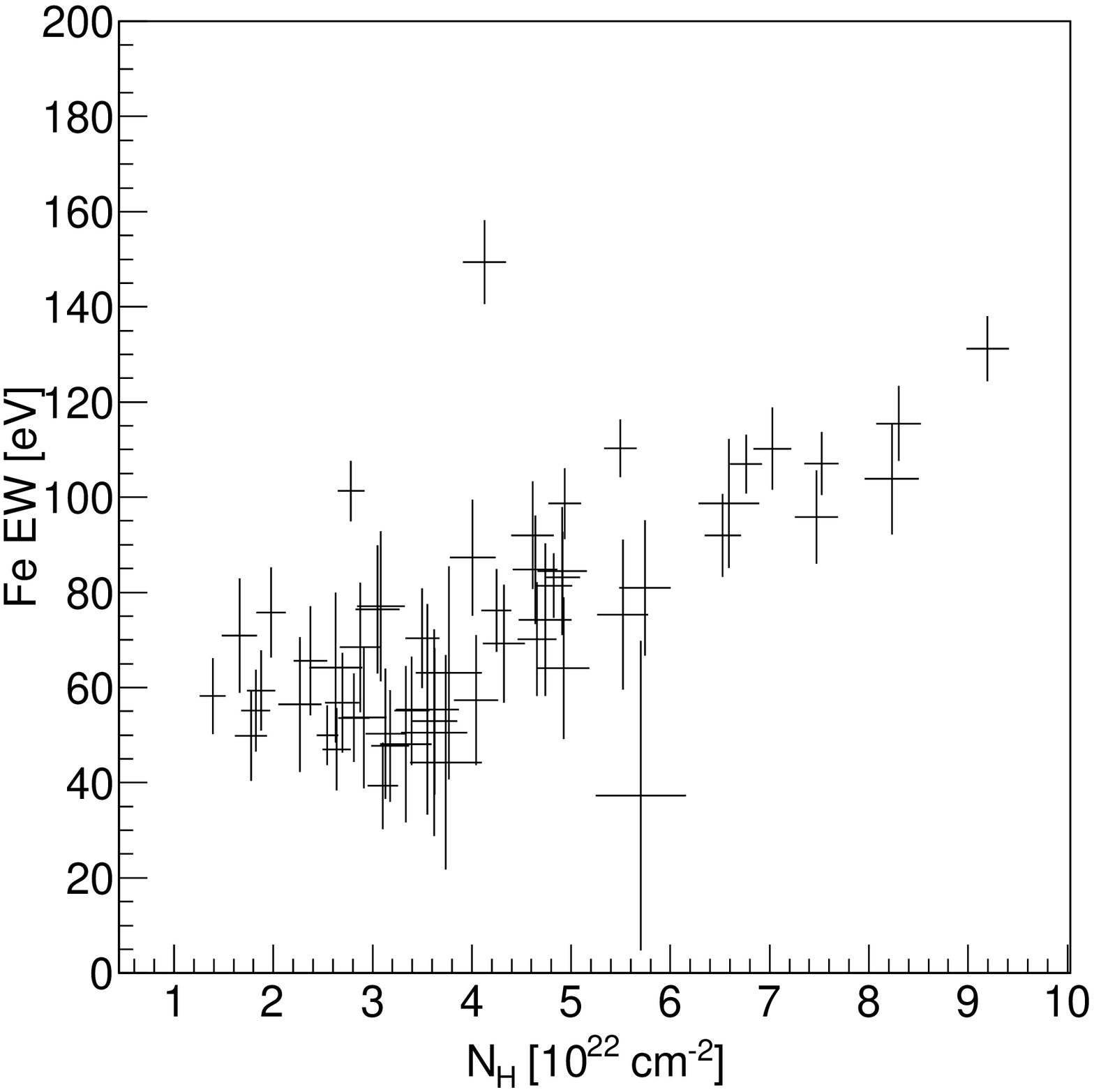}
\caption{Relation between the equivalent width of iron K$\alpha$ line
  and $N_\mathrm{H}$.}
\label{fig:plot_corr_xis_ew_nh}
\end{center}
\end{figure}

We also checked correlations between spectral parameters obtained over the whole observation.
Figures~\ref{fig:plot_corr_xis_flux_nh} and \ref{fig:plot_corr_xis_flux_ew} show the obtained relations between the X-ray flux and  two parameters derived though the  spectral fitting ($N_\mathrm{H}$ and the iron K$\alpha$ line equivalent width).
Both the X-ray absorption measured by $N_\mathrm{H}$ and the equivalent width of the iron K$\alpha$ line do not show any significant correlations with the X-ray flux.
By contrast, there is a positive correlation between $N_\mathrm{H}$ and the equivalent width of iron K$\alpha$ line, as shown in Figure~\ref{fig:plot_corr_xis_ew_nh}.

\begin{table*}[p]
\begin{minipage}{\textwidth}
\begin{center}
\caption{Results of spectral fit of the XIS data to power-law models}\label{table:fit_xis_pl}
{\scriptsize
\begin{tabular}{rrcccccccc}
  \hline\hline
  ID & Time & $N_\mathrm{H}$                     & $\tau_\mathrm{edge}$ & Photon index & Power-law                        & Fe norm.$^\mathrm{b}$                      & Fe EW$^\mathrm{c}$ & Flux (2--10 keV) & $\chi^2/\mathrm{dof}$ \\
       & [s]    & [$10^{22}\ \mathrm{cm}^{-2}$] &                                     & $\Gamma$     & norm. $A$$^\mathrm{a}$ & [$10^{-3}\ \mathrm{ph~cm^{-2}s^{-1}}$] & [eV] & [$10^{-9}\ \mathrm{erg~cm^{-2}s^{-1}}$]& \\
  \hline
     0 & 1134 & $4.83\pm0.18$  & $0.15\pm0.02$  & $0.73\pm0.03$  & $0.231\pm0.014$  & $4.86\pm0.42$  & $81\pm7$  & $3.742\pm0.017$  & $1.12$  \\
   1 & 3402 & $3.10\pm0.16$  & $0.07\pm0.02$  & $1.13\pm0.03$  & $0.305\pm0.017$  & $1.46\pm0.33$  & $39\pm9$  & $2.590\pm0.010$  & $0.99$  \\
   2 & 5671 & $2.54\pm0.11$  & $0.03\pm0.01$  & $1.09\pm0.02$  & $0.367\pm0.014$  & $2.41\pm0.29$  & $50\pm6$  & $3.484\pm0.010$  & $0.99$  \\
   4 & 10207 & $2.91\pm0.23$  & $0.06\pm0.03$  & $1.29\pm0.04$  & $0.121\pm0.010$  & $0.60\pm0.16$  & $54\pm15$  & $0.797\pm0.005$  & $1.05$  \\
   5 & 12475 & $3.33\pm0.26$  & $0.11\pm0.04$  & $1.30\pm0.05$  & $0.168\pm0.015$  & $0.72\pm0.24$  & $48\pm16$  & $1.051\pm0.008$  & $0.97$  \\
   6 & 14744 & $2.43\pm0.25$  & $0.05\pm0.03$  & $1.14\pm0.05$  & $0.230\pm0.020$  & $1.52\pm0.41$  & $55\pm15$  & $2.013\pm0.015$  & $1.03$  \\
   7 & 17012 & $2.64\pm0.14$  & $0.07\pm0.02$  & $1.20\pm0.03$  & $0.253\pm0.013$  & $1.28\pm0.23$  & $47\pm9$  & $1.961\pm0.008$  & $0.93$  \\
   8 & 19280 & $2.43\pm0.26$  & $0.03\pm0.04$  & $1.23\pm0.05$  & $0.308\pm0.029$  & $1.80\pm0.48$  & $57\pm16$  & $2.321\pm0.019$  & $1.00$  \\
   9 & 21548 & $1.88\pm0.14$  & $0.04\pm0.02$  & $1.14\pm0.03$  & $0.304\pm0.016$  & $2.19\pm0.31$  & $59\pm8$  & $2.758\pm0.011$  & $0.96$  \\
  10 & 23817 & $1.82\pm0.15$  & $0.09\pm0.02$  & $1.06\pm0.03$  & $0.272\pm0.014$  & $2.08\pm0.32$  & $55\pm9$  & $2.782\pm0.012$  & $0.96$  \\
  11 & 26085 & $2.63\pm0.27$  & $0.08\pm0.04$  & $1.17\pm0.05$  & $0.228\pm0.022$  & $1.68\pm0.41$  & $64\pm16$  & $1.879\pm0.017$  & $0.99$  \\
  12 & 28353 & $3.13\pm0.20$  & $0.10\pm0.03$  & $1.30\pm0.04$  & $0.154\pm0.011$  & $0.68\pm0.18$  & $50\pm14$  & $0.965\pm0.006$  & $1.00$  \\
  13 & 30621 & $3.74\pm0.36$  & $0.12\pm0.05$  & $1.35\pm0.07$  & $0.115\pm0.015$  & $0.41\pm0.21$  & $44\pm23$  & $0.638\pm0.008$  & $1.05$  \\
  14 & 32890 & $3.18\pm0.19$  & $0.07\pm0.03$  & $1.25\pm0.04$  & $0.251\pm0.017$  & $1.17\pm0.28$  & $48\pm12$  & $1.731\pm0.009$  & $0.96$  \\
  15 & 35158 & $2.69\pm0.17$  & $0.08\pm0.02$  & $1.25\pm0.03$  & $0.248\pm0.015$  & $1.37\pm0.25$  & $57\pm10$  & $1.744\pm0.008$  & $1.00$  \\
  16 & 37426 & $3.77\pm0.33$  & $0.04\pm0.05$  & $1.55\pm0.07$  & $0.119\pm0.014$  & $0.42\pm0.15$  & $63\pm22$  & $0.478\pm0.005$  & $0.96$  \\
  \rowcolor{tablegray} 17 & 39694 & $5.70\pm0.45$  & $0.06\pm0.07$  & $2.22\pm0.09$  & $0.157\pm0.026$  & $0.09\pm0.08$  & $37\pm33$  & $0.183\pm0.003$  & $0.96$  \\
  18 & 41963 & $3.62\pm0.33$  & $0.13\pm0.05$  & $1.37\pm0.07$  & $0.124\pm0.015$  & $0.49\pm0.21$  & $51\pm22$  & $0.677\pm0.007$  & $0.89$  \\
  19 & 44231 & $4.04\pm0.22$  & $0.07\pm0.03$  & $1.35\pm0.04$  & $0.157\pm0.012$  & $0.74\pm0.18$  & $57\pm14$  & $0.879\pm0.006$  & $0.98$  \\
  20 & 46499 & $2.87\pm0.20$  & $0.10\pm0.03$  & $1.26\pm0.04$  & $0.161\pm0.012$  & $1.06\pm0.20$  & $68\pm14$  & $1.110\pm0.006$  & $0.91$  \\
  21 & 48768 & $3.05\pm0.22$  & $0.09\pm0.03$  & $1.20\pm0.04$  & $0.164\pm0.013$  & $1.35\pm0.24$  & $76\pm13$  & $1.252\pm0.008$  & $0.96$  \\
  22 & 51036 & $2.81\pm0.16$  & $0.07\pm0.02$  & $1.18\pm0.03$  & $0.199\pm0.011$  & $1.20\pm0.21$  & $54\pm9$  & $1.594\pm0.007$  & $1.04$  \\
  23 & 53304 & $1.77\pm0.16$  & $0.02\pm0.02$  & $1.12\pm0.03$  & $0.248\pm0.014$  & $1.56\pm0.31$  & $50\pm10$  & $2.359\pm0.011$  & $1.08$  \\
  24 & 55572 & $1.39\pm0.13$  & $0.03\pm0.02$  & $1.12\pm0.03$  & $0.260\pm0.012$  & $1.89\pm0.26$  & $58\pm8$  & $2.498\pm0.010$  & $0.96$  \\
  25 & 57841 & $2.37\pm0.17$  & $0.08\pm0.02$  & $1.32\pm0.03$  & $0.215\pm0.013$  & $1.21\pm0.20$  & $66\pm11$  & $1.373\pm0.006$  & $0.98$  \\
  26 & 60109 & $1.66\pm0.18$  & $0.03\pm0.03$  & $1.26\pm0.04$  & $0.260\pm0.017$  & $1.78\pm0.29$  & $71\pm12$  & $1.949\pm0.010$  & $1.02$  \\
  27 & 62377 & $1.98\pm0.15$  & $0.04\pm0.02$  & $1.24\pm0.03$  & $0.219\pm0.012$  & $1.66\pm0.21$  & $76\pm9$  & $1.649\pm0.007$  & $0.98$  \\
  28 & 64645 & $2.27\pm0.22$  & $0.04\pm0.03$  & $1.29\pm0.04$  & $0.163\pm0.013$  & $0.84\pm0.21$  & $56\pm14$  & $1.109\pm0.007$  & $0.98$  \\
  29 & 66914 & $3.08\pm0.24$  & $0.02\pm0.03$  & $1.48\pm0.05$  & $0.169\pm0.015$  & $0.84\pm0.18$  & $77\pm16$  & $0.808\pm0.006$  & $1.00$  \\
  30 & 69182 & $2.16\pm0.27$  & $0.08\pm0.04$  & $1.29\pm0.05$  & $0.114\pm0.011$  & $0.74\pm0.19$  & $71\pm19$  & $0.783\pm0.007$  & $1.05$  \\
  31 & 71450 & $3.55\pm0.32$  & $0.00\pm0.09$  & $1.49\pm0.06$  & $0.179\pm0.020$  & $0.62\pm0.23$  & $55\pm22$  & $0.809\pm0.010$  & $0.96$  \\
  32 & 73718 & $3.62\pm0.23$  & $0.01\pm0.04$  & $1.54\pm0.05$  & $0.181\pm0.015$  & $0.55\pm0.16$  & $53\pm15$  & $0.750\pm0.006$  & $0.97$  \\
  34 & 78255 & $5.52\pm0.26$  & $0.05\pm0.03$  & $1.43\pm0.05$  & $0.197\pm0.017$  & $1.04\pm0.21$  & $75\pm16$  & $0.893\pm0.006$  & $1.01$  \\
  35 & 80523 & $3.50\pm0.17$  & $0.04\pm0.02$  & $1.30\pm0.03$  & $0.252\pm0.015$  & $1.57\pm0.23$  & $70\pm10$  & $1.579\pm0.007$  & $1.01$  \\
  37 & 85060 & $3.39\pm0.18$  & $0.04\pm0.02$  & $1.32\pm0.03$  & $0.207\pm0.013$  & $0.99\pm0.20$  & $55\pm11$  & $1.274\pm0.006$  & $1.07$  \\
  38 & 87328 & $4.92\pm0.26$  & $0.08\pm0.03$  & $1.29\pm0.05$  & $0.212\pm0.019$  & $1.23\pm0.28$  & $64\pm15$  & $1.249\pm0.009$  & $1.05$  \\
  39 & 89596 & $4.74\pm0.27$  & $0.06\pm0.03$  & $1.21\pm0.05$  & $0.154\pm0.014$  & $1.22\pm0.25$  & $74\pm16$  & $1.074\pm0.008$  & $0.96$  \\
  40 & 91864 & $4.66\pm0.20$  & $0.07\pm0.03$  & $1.26\pm0.04$  & $0.182\pm0.012$  & $1.22\pm0.20$  & $70\pm12$  & $1.150\pm0.006$  & $0.99$  \\
  42 & 96401 & $4.32\pm0.21$  & $0.10\pm0.03$  & $1.19\pm0.04$  & $0.148\pm0.011$  & $1.12\pm0.20$  & $69\pm12$  & $1.072\pm0.006$  & $0.98$  \\
  43 & 98669 & $5.74\pm0.26$  & $0.12\pm0.03$  & $1.32\pm0.05$  & $0.172\pm0.015$  & $1.21\pm0.21$  & $81\pm14$  & $0.935\pm0.007$  & $0.98$  \\
  44 & 100937 & $4.71\pm0.32$  & $0.09\pm0.04$  & $1.14\pm0.06$  & $0.205\pm0.022$  & $1.53\pm0.42$  & $62\pm19$  & $1.602\pm0.015$  & $1.02$  \\
  45 & 103206 & $4.25\pm0.15$  & $0.09\pm0.02$  & $1.15\pm0.03$  & $0.249\pm0.013$  & $2.26\pm0.24$  & $76\pm9$  & $1.962\pm0.008$  & $0.98$  \\
  46 & 105474 & $4.01\pm0.23$  & $0.09\pm0.03$  & $1.14\pm0.04$  & $0.222\pm0.017$  & $2.35\pm0.33$  & $87\pm12$  & $1.805\pm0.011$  & $0.92$  \\
  \rowcolor{tablegray} 47 & 107742 & $8.23\pm0.27$  & $0.16\pm0.03$  & $0.97\pm0.04$  & $0.149\pm0.012$  & $2.54\pm0.28$  & $104\pm12$  & $1.372\pm0.009$  & $1.05$  \\
  \rowcolor{tablegray} 48 & 110010 & $4.13\pm0.22$  & $0.23\pm0.02$  & $0.46\pm0.04$  & $0.084\pm0.006$  & $5.36\pm0.31$  & $149\pm9$  & $2.272\pm0.012$  & $1.07$  \\
  49 & 112279 & $6.59\pm0.31$  & $0.10\pm0.03$  & $1.01\pm0.05$  & $0.211\pm0.021$  & $3.18\pm0.43$  & $99\pm14$  & $1.939\pm0.016$  & $0.98$  \\
  50 & 114547 & $6.53\pm0.18$  & $0.08\pm0.02$  & $1.14\pm0.03$  & $0.221\pm0.013$  & $2.47\pm0.23$  & $92\pm9$  & $1.628\pm0.008$  & $1.03$  \\
  51 & 116815 & $4.91\pm0.25$  & $0.13\pm0.03$  & $1.16\pm0.05$  & $0.161\pm0.013$  & $1.58\pm0.25$  & $84\pm13$  & $1.195\pm0.008$  & $1.05$  \\
  52 & 119083 & $4.64\pm0.22$  & $0.15\pm0.03$  & $1.05\pm0.04$  & $0.143\pm0.011$  & $1.74\pm0.24$  & $85\pm11$  & $1.313\pm0.007$  & $0.96$  \\
  53 & 121352 & $4.91\pm0.18$  & $0.14\pm0.02$  & $1.16\pm0.03$  & $0.209\pm0.013$  & $2.01\pm0.23$  & $83\pm10$  & $1.542\pm0.007$  & $0.99$  \\
  54 & 123620 & $4.61\pm0.21$  & $0.15\pm0.03$  & $1.10\pm0.04$  & $0.189\pm0.014$  & $2.25\pm0.27$  & $92\pm11$  & $1.575\pm0.009$  & $0.95$  \\
  55 & 125888 & $7.03\pm0.19$  & $0.11\pm0.02$  & $1.05\pm0.03$  & $0.196\pm0.012$  & $3.04\pm0.23$  & $110\pm9$  & $1.633\pm0.007$  & $1.04$  \\
  56 & 128156 & $7.47\pm0.22$  & $0.11\pm0.02$  & $1.08\pm0.04$  & $0.213\pm0.015$  & $2.76\pm0.27$  & $96\pm10$  & $1.676\pm0.009$  & $0.96$  \\
  57 & 130425 & $4.93\pm0.16$  & $0.17\pm0.02$  & $0.93\pm0.03$  & $0.201\pm0.011$  & $3.51\pm0.26$  & $99\pm7$  & $2.227\pm0.009$  & $1.03$  \\
  58 & 132693 & $6.76\pm0.16$  & $0.17\pm0.02$  & $0.82\pm0.03$  & $0.193\pm0.010$  & $4.52\pm0.26$  & $107\pm6$  & $2.471\pm0.009$  & $0.92$  \\
  59 & 134961 & $8.30\pm0.22$  & $0.19\pm0.02$  & $0.79\pm0.04$  & $0.179\pm0.012$  & $4.80\pm0.32$  & $116\pm8$  & $2.306\pm0.011$  & $1.02$  \\
  60 & 137229 & $7.52\pm0.17$  & $0.19\pm0.02$  & $0.76\pm0.03$  & $0.166\pm0.009$  & $4.38\pm0.26$  & $107\pm7$  & $2.325\pm0.009$  & $0.90$  \\
  61 & 139498 & $9.20\pm0.21$  & $0.25\pm0.02$  & $0.63\pm0.03$  & $0.135\pm0.008$  & $5.44\pm0.27$  & $131\pm7$  & $2.220\pm0.009$  & $1.09$  \\
  62 & 141766 & $5.50\pm0.16$  & $0.14\pm0.02$  & $0.85\pm0.03$  & $0.256\pm0.013$  & $5.82\pm0.32$  & $110\pm6$  & $3.249\pm0.012$  & $1.00$  \\
  63 & 144034 & $2.78\pm0.13$  & $0.14\pm0.02$  & $0.87\pm0.03$  & $0.224\pm0.010$  & $4.52\pm0.28$  & $101\pm6$  & $3.069\pm0.011$  & $0.93$  \\

  \hline
\end{tabular}
}
\end{center}
\begin{flushleft} {\scriptsize All errors represent 1$\sigma$
    uncertainties. a) normalization of the power law in units of
    $\mathrm{photons~keV^{-1}cm^{-2}s^{-1}}$ at 1 keV. b)
    normalization of Fe K$\alpha$ at 6.4 keV. c) equivalent width (EW) of
    Fe K$\alpha$ at 6.4 keV to underlying continuum.
    The data set referred in the text via the ID numbers are marked with gray color.}
\end{flushleft}
\end{minipage}
\end{table*}

\section{XIS/HXD Wide-Band Analysis}\label{sec:analysis_xishxd}

In this section we present the broad-band analysis of the data obtained with both XIS and HXD onboard {\it Suzaku} from the Vela X-1 pulsar.
The key difference of this analysis to the one conducted in the previous section is that the spectral fitting was performed for a broader range of energies.
Namely, the data obtained in the soft (2.5--10~keV) and hard X-ray (15--90~keV) energy bands were used simultaneously for the spectral fitting.
The soft X-ray data were the same as in the previous section and the hard X-ray data were obtained with the HXD.

\subsection{Light Curves}

\begin{figure}[tbp]
\begin{center}
\includegraphics[width=9cm]{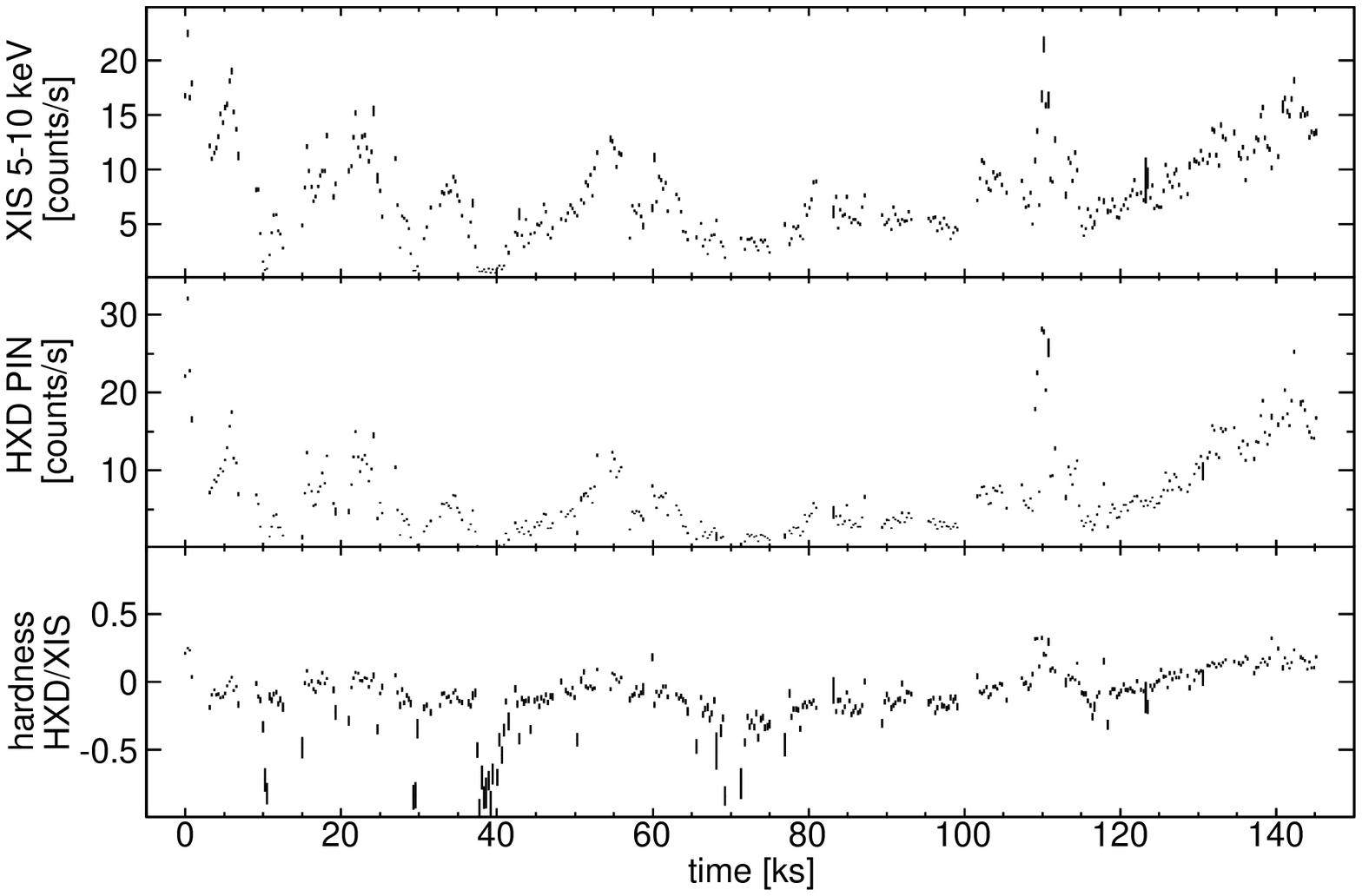}
\caption{Light curves of XIS 0 (5--10 keV) and HXD-PIN above $\sim 12$
  keV. A hardness ratio between the different detectors are also
  presented. The time bin size is 284 s, the same as the spin period
  of the neutron star. The HXD-PIN light curve are
  background-subtracted and dead-time-corrected.}
\label{fig:hxd_lc_284s}
\end{center}
\end{figure}

Figure~\ref{fig:hxd_lc_284s} shows light curves of the HXD-PIN data above $\sim$12 keV together with the XIS 0 data (5--10 keV).
The curves were extracted from the two detectors with a time bin width of 284 s, which corresponds to the spin period (the same as Figure~\ref{fig:xis_lc_284s}).
The HXD-PIN light curves are background-subtracted and dead-time-corrected though the non-Xray background of the HXD-PIN detector was $\sim$0.5 $\rm counts\, s^{-1}$, which is not significant when compared to such a bright source as Vela X-1.

While the temporal behavior in the two bands is very similar on the whole, the hardness ratio increases with the luminosity.
The low states (low-luminosity states with spectral softening) found in the soft band are also clearly seen in the wide-band analysis (e.g. see a strong dip in the hardness ratio around $t\sim40$~ks).
The flare at $t=110$ ks also showed a significant hardening.

\subsection{Spectral Model}

Since there is no well-established physical model of the radiation from accreting X-ray pulsars, we introduce a phenomenological spectral model to quantify the {\it Suzaku} wide-band data.
For the detailed time-resolved analysis, we constructed a spectral model template by using the time-averaged spectrum of Vela X-1 obtained with the XIS and HXD.
The spectra for the simultaneous wide-band modeling were extracted from data of XIS 0, XIS 3, HXD-PIN, and HXD-GSO with common time intervals for these detectors.

There are three well-known phenomenological spectral models to describe a power law with quasi-exponential cutoff at high energy, i.e., a spectrum typical for X-ray pulsars
\citep[e.g.][]{Kreykenbohm:1999}.
A power law with a high-energy cut is one of them \citep{White:1983}, written as
\begin{equation}\label{eq:model_pl_highecut}
\frac{dN}{dE} = \left\{ \begin{array}{lc}
AE^{-\Gamma} & (E\le E_c) \\
AE^{-\Gamma}\exp\left(-\dfrac{E-E_c}{E_f}\right) & (E>E_c)
\end{array}\right.
\end{equation}
where $E_c$ and $E_f$ are referred to as ``cutoff energy'' and  ``folding energy'', respectively.
This model converges to the ``normal'' cutoff power law (power law with an exponential cut) if $E_c \to 0$.
Although this model has been often used for spectral modeling of the emission of accreting neutron stars, its derivative is discontinuous at $E=E_c$, and this could introduce artificial features in the continuum.

The second one is a power law with the so-called Fermi-Dirac cutoff \citep{Tanaka:1986, Kreykenbohm:1999}, which renders  a spectrum with a smooth turnover:
\begin{equation}\label{eq:model_pl_fdcut}
\frac{dN}{dE} = AE^{-\Gamma}\frac{1}{\exp\left(\dfrac{E-E_c}{E_f}\right)+1}.
\end{equation}
This function is smooth at any point, and nicely fits a spectrum with a cutoff.
Due to historical reasons, the parameters of this model have the same names as those of the high-energy cut model.
However, the actual meaning of these parameters is very different, therefore they cannot be compared directly.

The third is the negative and positive exponential cutoff power law (NPEX) model \citep{Mihara:1995, Makishima:1999},
\begin{gather}
\frac{dN}{dE} = (A_1E^{-\Gamma_1}+A_2E^{+\Gamma_2})\exp\left(-\dfrac{E}{E_f}\right),\\
\Gamma_1 > 0, \quad \Gamma_2 > 0.
\end{gather}
This is a combination of negative and positive power laws with an exponential cutoff by a common folding energy.
The combination of the two power laws with largely different indices provide good fits for a wide variety of spectral shapes.
If the positive power-law index is
$\Gamma_2=2$ as
\begin{equation}\label{eq:model_npex}
\frac{dN}{dE} = (A_1E^{-\Gamma}+A_2E^{2})\exp\left(-\dfrac{E}{E_f}\right),
\end{equation}
the function is known to be a good approximation of the unsaturated Comptonization spectrum \citep{Sunyaev:1980}.
\citet{Mihara:1995} found that X-ray pulsar spectra obtained with the {\it Ginga} satellite are well described by this model \citep[see also][]{Makishima:1999}.
The NPEX model approaches to a normal cutoff power law when $A_2\to 0$.

Examining these three models, we select the NPEX function (Eq.~\ref{eq:model_npex}) to characterize the spectrum and to extract physical information.
The high-energy cut model (Eq.~\ref{eq:model_pl_highecut}) is excluded because of the discontinuity of the derivative.
Although the Fermi-Dirac cutoff model (Eq.~\ref{eq:model_pl_fdcut}) well describes the spectral shape, results of preliminary fitting to the obtained data have shown that there is strong coupling between the power-law normalization $A$ and the cutoff energy $E_\mathrm{c}$ near $E_\mathrm{c}\sim 0$, which generates large uncertainties of these parameters.
The NPEX function is smooth and easy to analyze because it simply consists of two power laws with a common exponential cutoff.
Furthermore, the model parameters allow a tentative physical interpretation in the context of thermal Comptonization \citep{Sunyaev:1980, Makishima:1999}.

Many X-ray pulsars display absorption-like features in the hard X-ray continuum called cyclotron resonance scattering features (CRSFs).
They are a result of scattering via cyclotron resonance of electrons whose energy is quantized to the Landau levels in the strong magnetic field ($B\sim 10^{12}$~G).
The CRSFs have complex structures which are composed of absorption and emission via cyclotron radiation from electrons in hot magnetized plasma \citep{Araya:1999, ArayaGochez:2000, Schonherr:2007}.
It is, however, difficult to include these complex models of the CRSF in the spectral fitting with the detector response.
We therefore adopt a widely-used phenomenological model for cyclotron absorptions \citep[e.g.][]{Mihara:1990, Makishima:1999, Kreykenbohm:2008}.
The effective optical depth due to the cyclotron absorptions can be written as
\begin{equation}\label{eq:model_cyclotron_absorption}
\begin{split}
\tau_\mathrm{cycl}(E)=&D_1\frac{(W_1 E/E_\mathrm{cycl})^2}{(E-E_\mathrm{cycl})^2+W_1{}^2} \\
&+D_2\frac{(W_2 E/2E_\mathrm{cycl})^2}{(E-2E_\mathrm{cycl})^2+W_2{}^2},
\end{split}
\end{equation}
where $E_\mathrm{cycl}$ is the cyclotron energy, $W_1$, $W_2$, $D_1$, $D_2$ denotes widths and depths for the fundamental (subscript 1) and
the first (subscript 2) harmonic resonances.
The effect of the absorption in the spectral fitting is then included by the factor $\exp(-\tau_\mathrm{cycl})$.

Combining the continuum model with the effects of absorption and emission from the circumstellar medium such as the stellar wind which is described by Equation~(\ref{eq:spec_model_xis}), we introduce a spectral model, based on the NPEX function, for the wide-band fitting as the following:
\begin{equation}\label{eq:full_spec_model_npex}
\begin{split}
\frac{dN}{dE}=&\exp(-N_\text{H}\sigma_\text{abs}(E))f_\text{edge}(E)\\
&\times \left[(A_1E^{-\Gamma}+A_2E^{2})\exp\left(-\dfrac{E}{E_f}\right)+F_\text{Fe}(E)\right] \\
&\times \exp(-\tau_\mathrm{cycl}(E)).
\end{split}
\end{equation}
Since the model described by Equation (\ref{eq:full_spec_model_npex}) is very complex and the $\chi^2$ minimization might converge to an undesired local minimum, we preliminarily fitted the global structure of the spectrum, and then set the initial values of the parameters of subsequent fits to the parameters obtained in the previous fit.

\begin{figure*}[htbp]
\begin{center}
\includegraphics[width=13.5cm]{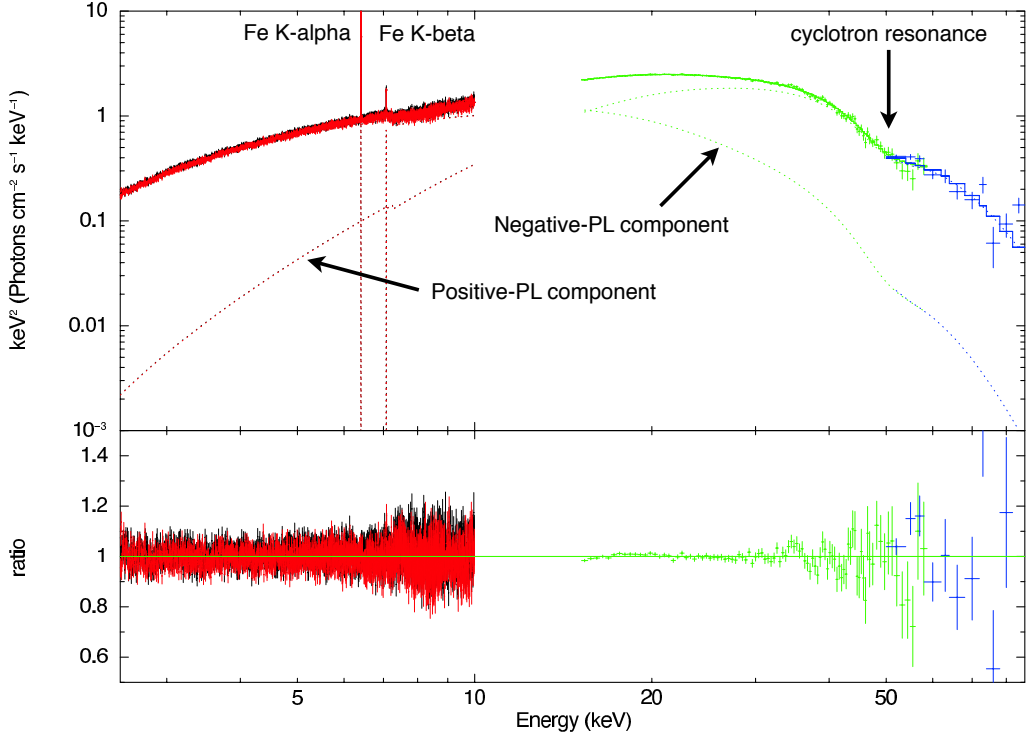}
\caption{Time-averaged wide-band X-ray spectrum of Vela X-1 and the
  NPEX model fitted to the data. The data of XIS 0 (black), XIS 3
  (red), HXD-PIN (green), and HXD-GSO (blue) are shown. The top panel
  shows unfolded $\nu F_\nu$ spectrum with the detector response (top)
  and the bottom shows the data-to-model ratio. The dashed lines show 
  the two power-law components of the NPEX function. The model includes the
  CRSFs.}
\label{fig:spec_taverage_npex}
\end{center}
\end{figure*}

The fitting result of the time-averaged spectrum to the NPEX model is shown in Table~\ref{table:fit_taverage_npex}.
First we tried to fit the spectrum to the model without the CRSFs, but the data-to-model ratio showed a dip at $E\sim 50$ keV, which is probably a CRSF at 50--55 keV that has been reported in many observations \citep[e.g.][]{Orlandini:1998, Makishima:1999, Kreykenbohm:1999, Kreykenbohm:2002}.
In addition, some observations reported an existence of a CRSF at 25 keV \citep{Makishima:1999, Kreykenbohm:2002} while the observation with {\it Beppo-SAX} could not find a strong evidence of the feature around $E\sim 25$ keV \citep{Orlandini:1998}.
Our preliminary fit also showed a shallow dip at 25 keV.
Considering these facts, we added two CRSFs with the cyclotron energy $E_\mathrm{cycl}\simeq 25$ keV (equal to the fundamental line) to the preliminary model.
The final results are shown in Figure~\ref{fig:spec_taverage_npex}.
The model also contains the Fe K$\beta$ line at $E=7.1$ keV, but the significance of this line is not high.

There are still residuals between the model and the data, probably, because the adopted spectral model, which is a combination of the NPEX function and the simple absorption-like functions describing the CRSFs. Indeed it is obvious that this simple model is not enough to represent the complicated integration of the neutron star emission over the long time and over the different emission regions.
 
We used a spectral model with two CRSFs, the fundamental line at $\sim$25 keV and the first harmonic line at $\sim$50 keV, for fitting the data.
While the CRSF at $E=50$ keV is clearly seen in the spectrum, we do not find any strong evidence of the CRSF at 25 keV for the time-averaged spectrum.
Thus, the shallow dip at 25 keV, which was attributed to a CRSF at the preliminary analysis, is likely to related a discrepancy between the simple continuum model and the complicated neutron star emission.
However, we note that \citet{Kreykenbohm:2002} showed the existence of the fundamental line at 25 keV based on spin-phase-resolved spectroscopy, since this feature is much weaker than the CRSF at 50 keV and its visibility seems to depends on the pulsar phase.
Thus, to confirm or discard its existence it would be important to perform the phase-resolved analysis, but such a study requires an additional dedicated analysis, and will be discussed elsewhere.

\begin{table*}[htdp]
  \caption{Results of spectral fit of the time-averaged wide-band spectrum to the NPEX model}
  \label{table:fit_taverage_npex}
\begin{center}
\begin{tabular}{cc}
\hline\hline
Parameter & Value \\
\hline
$N_\mathrm{H}$ [$10^{22}$ cm$^{-2}$] & $2.495\pm 0.043$ \\
$E_\text{edge}$ [keV] & $7.255\pm 0.017$ \\
$\tau_\text{edge}$ & $0.095\pm 0.004$ \\
\hline
Negative power-law photon index $\Gamma$ & $0.385\pm 0.016$ \\
Negative power-law normalization $A_1$ [photons s$^{-1}$cm$^{-2}$ at 1 keV] & $0.1076\pm 0.0019$ \\
Positive power-law normalization $A_2$ [$10^{-3}$ photons s$^{-1}$cm$^{-2}$ at 1 keV] & $0.152\pm 0.013$ \\
Cutoff energy $E_f$ [keV] & $7.10\pm 0.21$ \\
\hline
Fe K$\alpha$ energy [keV] & $6.410\pm 0.008$ \\
Fe K$\alpha$ width $\sigma$ [keV] & $0.00\pm 0.0001$ \\
Fe K$\alpha$ normalization [$10^{-3}$ photons s$^{-1}$cm$^{-2}$] & $1.847\pm 0.025$ \\
Fe K$\beta$ energy [keV] & $7.08\pm 0.16$ \\
Fe K$\beta$ width $\sigma$ [keV] & $0.00\pm 0.04$ \\
Fe K$\beta$ normalization [$10^{-3}$ photons s$^{-1}$cm$^{-2}$] & $0.109 \pm 0.025$ \\
\hline
Fundamental cyclotron line depth $D_1$ & $0.036\pm 0.008$ \\
Fundamental cyclotron line energy $E_\mathrm{cycl}$ [keV] & $24.67\pm 0.25$ \\
Fundamental cyclotron line width $W_1$ [keV] & $9.0\pm 1.4$ \\
First harmonic cyclotron line depth $D_2$ & $0.528\pm 0.057$ \\
First harmonic cyclotron line energy $2E_\mathrm{cycl}$ [keV] & $49.34\pm 0.50$ (fixed to $2E_\mathrm{cycl}$) \\
First harmonic cyclotron line width $W_2$ & $9.0\pm 1.4$ (fixed to $W_1$) \\
\hline
$\chi^2/\mathrm{dof}$ & 4305/3845 \\
\hline
\end{tabular}
\end{center}

\end{table*}%

\subsection{Time-Resolved Spectra}\label{subsec:wide_band_time_resolved_spectrum}

Since the hard X-ray radiation of Vela X-1 displays a strong time variability, it is important to analyze wide-band X-ray spectra time-resolved on a short time scale.
We extracted wide-band spectra from the event data of XIS 0, XIS 3, XIS 1, and HXD-PIN for time bins of 2268 seconds (the same approach as for the XIS analysis described in \S\ref{subsec:xis_spectrum}).
In order to characterize the time evolution of the neutron star radiation, we performed spectral fitting of the obtained spectra to the NPEX model described by Equation~(\ref{eq:full_spec_model_npex}).  Due to the lower statistics of the time-resolved spectra as compared to the time-averaged one, the parameters related to the CRSF were fixed to the values obtained from the fit of the time-averaged spectrum.

\begin{figure*}[tbp]
\begin{center}
\includegraphics[width=16.5cm]{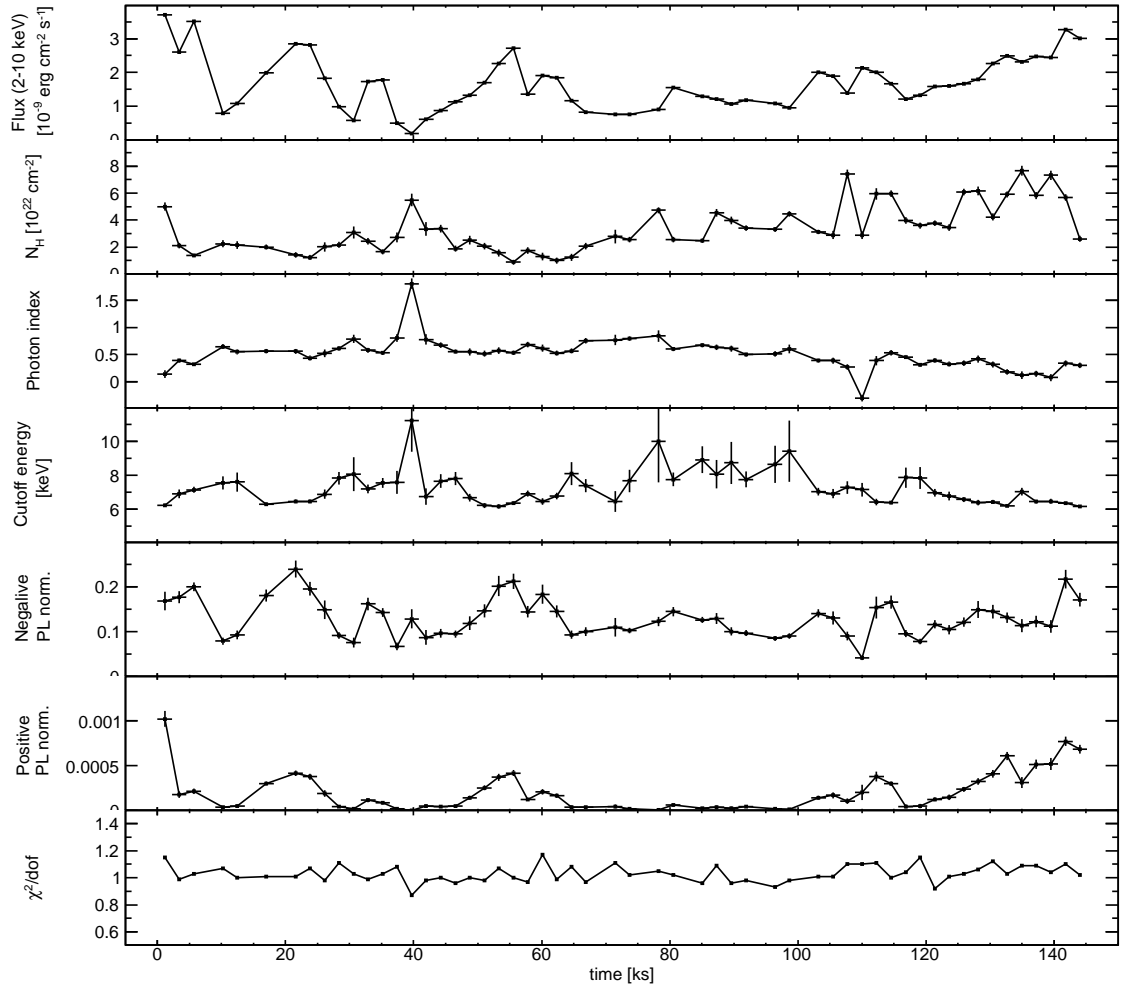}
\caption{Best-fit parameters of the XIS/HXD data to NPEX models as
  functions of time. The values are tabulated in
  Table~\ref{table:fit_xishxd_npex}.}
\label{fig:plot_fit_xishxd_npex}
\end{center}
\end{figure*}

The results of the fitting are tabulated in Table~\ref{table:fit_xishxd_npex} and plotted in Figure~\ref{fig:plot_fit_xishxd_npex}.
We excluded several spectra with exposure times shorter than 568 s from the fitting analysis to avoid large uncertainties due to low statistics and potential bias by the pulsar spin-phase.
Each obtained spectrum is well represented by the NPEX model.
The energy fluxes in 2--10 keV and $N_\mathrm{H}$ estimated here are consistent with the values obtained only from the soft X-ray data (Table~\ref{table:fit_xis_pl}).  Though the absolute values of the photon index of the NPEX model cannot be compared directly with the index of a single power law, the temporal behavior is very similar to the one obtained in the soft-X-ray analysis.

\begin{figure}[tbp]
\begin{center}
\includegraphics[width=7.0cm]{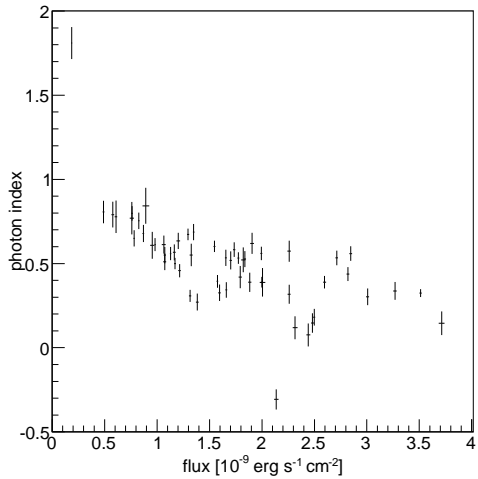}
\caption{Relation between the photon index $\Gamma$ of the NPEX
  function and the X-ray flux.}
\label{fig:plot_corr_gamma_flux}
\end{center}
\end{figure}

Together with the photon index $\Gamma$, the normalization of the positive power law $A_2$ and the cutoff energy $E_f$ characterize the hard X-ray radiation of the neutron star.  The photon index $\Gamma$ determines the slope of the flat power-law spectrum at lower energies.
Figure~\ref{fig:plot_corr_gamma_flux} shows the relation between the photon index and the flux, and the shown data imply that the spectrum becomes hard when the flux increases. 

\begin{figure}[tbp]
\begin{center}
\includegraphics[width=7.0cm]{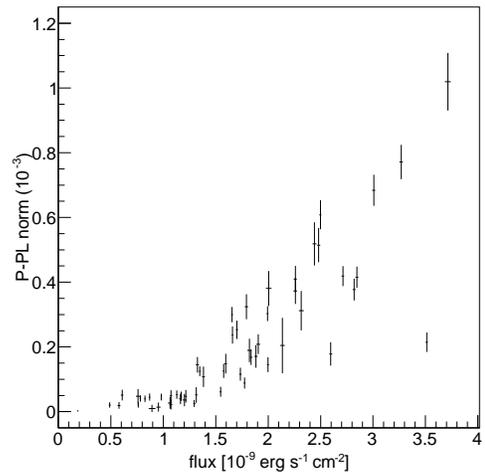}
\caption{Relation between the normalization of the positive power law
  $A_2$ and the X-ray flux.}
\label{fig:plot_corr_pnorm_flux}
\end{center}
\end{figure}

\begin{figure}[tbp]
\begin{center}
\includegraphics[width=7.0cm]{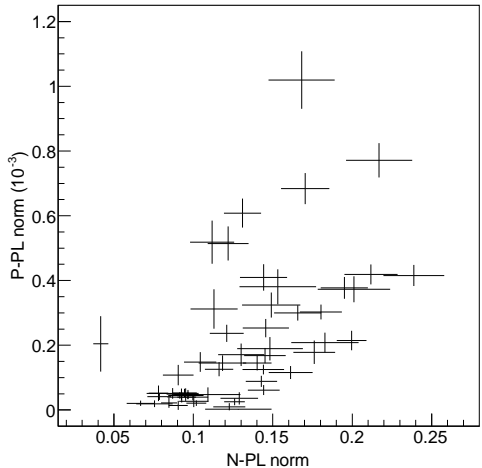}
\caption{Relation between the normalizations of the two power laws,
  $A_2$ and $A_1$.}
\label{fig:plot_corr_pnorm_nnorm}
\end{center}
\end{figure}

The two parameters of the NPEX model, $A_2$ and $E_f$ determine the spectral shape of the quasi-exponential cutoff at high energies, and therefore should reflect the physical conditions of the accreted plasma that radiates hard X-rays.
The relation between $A_2$ and the flux is shown in Figure~\ref{fig:plot_corr_pnorm_flux}.
If the spectral shape were  not changing with the total luminosity, $A_2$ would be exactly  proportional to the flux level.
But, interestingly, the positive power-law component is small and does not increase with flux if the flux is smaller than $\sim 1.5\times 10^{-9}\ \mathrm{erg\ cm^{-2} s^{-1}}$.  At higher fluxes, $A_2$ increases with the flux, like the ``normal'' negative power-law component.
This behavior of the positive power law is apparent in the correlation plot between $A_2$ and $A_1$, as shown in Figure~\ref{fig:plot_corr_pnorm_nnorm}.

\begin{figure}[tbp]
\begin{center}
\includegraphics[width=7.0cm]{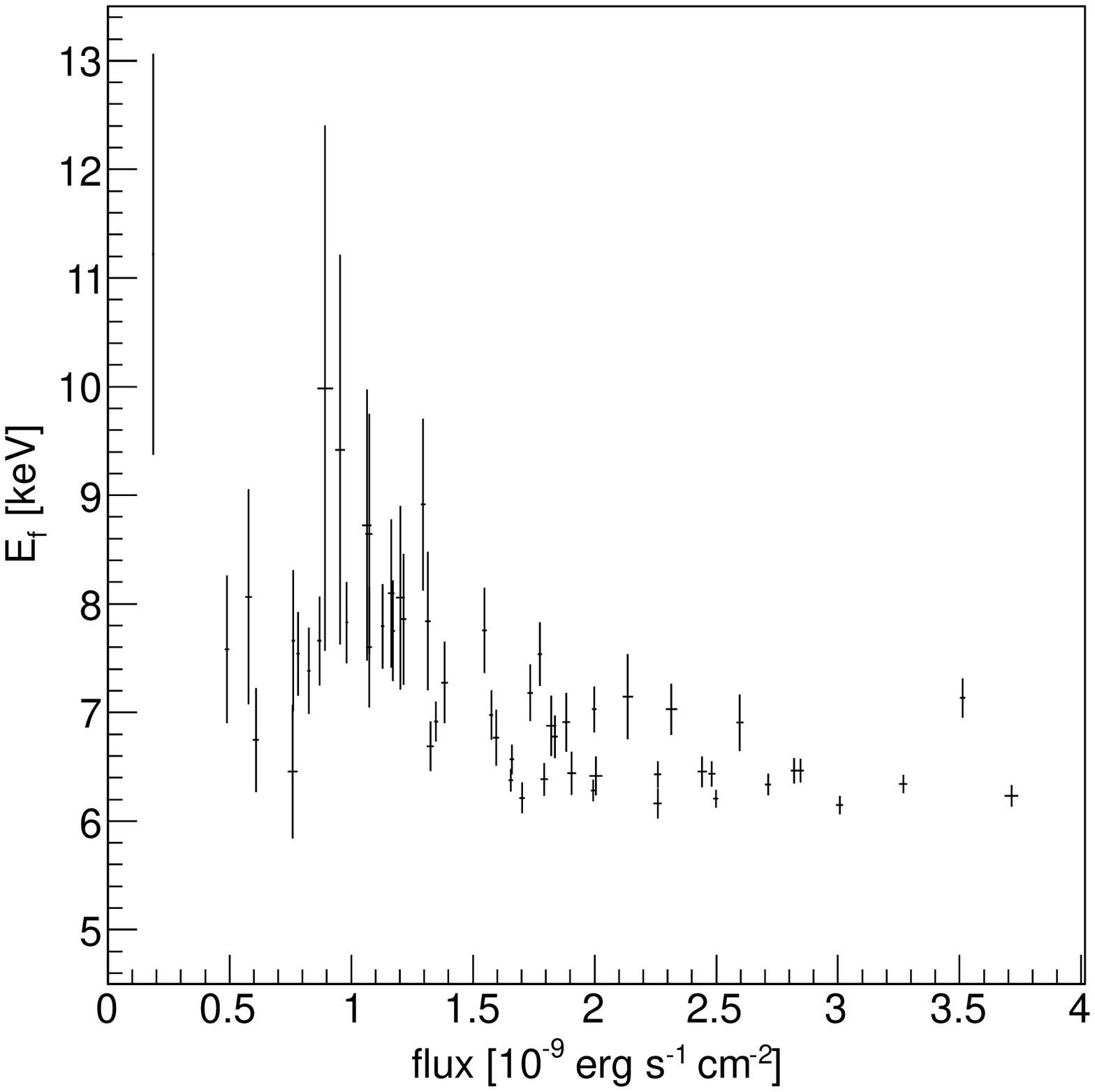}
\caption{Relation between the normalization of the cutoff energy $E_f$
  and the X-ray flux.}
\label{fig:plot_corr_ecut_flux}
\end{center}
\end{figure}

The cutoff energy $E_f$ determines the shape of the high-energy turnover.
The estimated values of $E_f$ range between 6 keV and 10 keV, and are concentrated around 6--7 keV.
The relation between the cutoff energy and the flux is shown in Figure~\ref{fig:plot_corr_ecut_flux}.
It seems to be weakly anticorrelated with the flux if the flux is below  $\sim 2\times 10^{-9}\ \mathrm{erg\ cm^{-2} s^{-1}}$.
For higher fluxes, the cutoff energy seems to be constant about $E_f\sim 6$ keV.

\begin{table*}[p]
\begin{minipage}{\textwidth}
\begin{center}
\caption{Results of spectral fit of the XIS/HXD data to NPEX models}\label{table:fit_xishxd_npex}
{\scriptsize
\begin{tabular}{rrcccccccc}
  \hline\hline
  ID & Time & $N_\mathrm{H}$                     & Photon index $^\mathrm{a}$ & Cutoff energy & Negative-PL                         & Positive-PL  & Flux (2--10 keV) & $\chi^2/\mathrm{dof}$ \\
      & [s]     & [$10^{22}\ \mathrm{cm}^{-2}$] & $\Gamma$                            & $E_f$ [keV]    & norm. $A_1$$^\mathrm{b}$ & norm $A_2$$^\mathrm{c}$       & [$10^{-9}\ \mathrm{erg~cm^{-2}s^{-1}}$]& \\
  \hline
     0 & 1134 & $5.00\pm0.31$  & $0.15\pm0.07$  & $6.23\pm0.10$  & $0.17\pm0.02$  & $1.02\pm0.09$  & $3.713\pm0.027$  & $1.15$  \\
   1 & 3402 & $2.10\pm0.20$  & $0.39\pm0.04$  & $6.91\pm0.26$  & $0.18\pm0.01$  & $0.18\pm0.04$  & $2.597\pm0.014$  & $0.99$  \\
   2 & 5671 & $1.37\pm0.12$  & $0.32\pm0.02$  & $7.13\pm0.18$  & $0.20\pm0.01$  & $0.21\pm0.03$  & $3.513\pm0.011$  & $1.03$  \\
   4 & 10207 & $2.24\pm0.25$  & $0.65\pm0.05$  & $7.54\pm0.38$  & $0.08\pm0.01$  & $0.04\pm0.01$  & $0.781\pm0.007$  & $1.07$  \\
   5 & 12475 & $2.14\pm0.27$  & $0.55\pm0.05$  & $7.60\pm0.56$  & $0.09\pm0.01$  & $0.05\pm0.02$  & $1.074\pm0.011$  & $1.00$  \\
   7 & 17012 & $1.99\pm0.18$  & $0.56\pm0.04$  & $6.28\pm0.10$  & $0.18\pm0.01$  & $0.30\pm0.02$  & $1.994\pm0.008$  & $1.01$  \\
   9 & 21548 & $1.41\pm0.19$  & $0.56\pm0.04$  & $6.46\pm0.11$  & $0.24\pm0.02$  & $0.42\pm0.03$  & $2.846\pm0.013$  & $1.01$  \\
  10 & 23817 & $1.22\pm0.18$  & $0.44\pm0.04$  & $6.46\pm0.12$  & $0.20\pm0.01$  & $0.38\pm0.03$  & $2.819\pm0.013$  & $1.07$  \\
  11 & 26085 & $2.01\pm0.35$  & $0.52\pm0.07$  & $6.88\pm0.28$  & $0.15\pm0.02$  & $0.19\pm0.04$  & $1.821\pm0.020$  & $0.98$  \\
  12 & 28353 & $2.16\pm0.21$  & $0.61\pm0.04$  & $7.83\pm0.37$  & $0.09\pm0.01$  & $0.05\pm0.01$  & $0.982\pm0.006$  & $1.11$  \\
  13 & 30621 & $3.07\pm0.43$  & $0.79\pm0.08$  & $8.07\pm0.99$  & $0.08\pm0.01$  & $0.02\pm0.01$  & $0.577\pm0.013$  & $1.03$  \\
  14 & 32890 & $2.43\pm0.23$  & $0.58\pm0.04$  & $7.18\pm0.26$  & $0.16\pm0.01$  & $0.12\pm0.02$  & $1.736\pm0.011$  & $0.99$  \\
  15 & 35158 & $1.65\pm0.18$  & $0.53\pm0.03$  & $7.54\pm0.29$  & $0.14\pm0.01$  & $0.09\pm0.02$  & $1.775\pm0.009$  & $1.03$  \\
  16 & 37426 & $2.70\pm0.36$  & $0.81\pm0.07$  & $7.58\pm0.68$  & $0.07\pm0.01$  & $0.02\pm0.01$  & $0.490\pm0.008$  & $1.08$  \\
  \rowcolor{tablegray} 17& 39694 & $5.48\pm0.46$  & $1.81\pm0.09$  & $11.22\pm1.85$  & $0.13\pm0.02$  & $0.00\pm0.00$  & $0.186\pm 0.005$  & $0.87$  \\
  18 & 41963 & $3.34\pm0.50$  & $0.78\pm0.10$  & $6.75\pm0.48$  & $0.09\pm0.02$  & $0.05\pm0.02$  & $0.608\pm0.013$  & $0.98$  \\
  19 & 44231 & $3.36\pm0.28$  & $0.68\pm0.05$  & $7.66\pm0.41$  & $0.10\pm0.01$  & $0.04\pm0.01$  & $0.870\pm0.008$  & $1.00$  \\
  20 & 46499 & $1.87\pm0.22$  & $0.56\pm0.04$  & $7.79\pm0.39$  & $0.09\pm0.01$  & $0.05\pm0.01$  & $1.129\pm0.007$  & $0.96$  \\
  21 & 48768 & $2.50\pm0.32$  & $0.55\pm0.07$  & $6.69\pm0.23$  & $0.12\pm0.02$  & $0.14\pm0.02$  & $1.325\pm0.014$  & $1.00$  \\
  22 & 51036 & $2.07\pm0.25$  & $0.52\pm0.05$  & $6.21\pm0.14$  & $0.15\pm0.01$  & $0.25\pm0.03$  & $1.703\pm0.011$  & $0.98$  \\
  23 & 53304 & $1.56\pm0.27$  & $0.57\pm0.06$  & $6.16\pm0.14$  & $0.20\pm0.02$  & $0.37\pm0.04$  & $2.259\pm0.017$  & $1.07$  \\
  24 & 55572 & $0.87\pm0.19$  & $0.53\pm0.04$  & $6.34\pm0.10$  & $0.21\pm0.02$  & $0.42\pm0.03$  & $2.713\pm0.012$  & $1.00$  \\
  25 & 57841 & $1.74\pm0.23$  & $0.69\pm0.05$  & $6.92\pm0.18$  & $0.14\pm0.01$  & $0.12\pm0.01$  & $1.349\pm0.008$  & $0.97$  \\
  26 & 60109 & $1.28\pm0.29$  & $0.62\pm0.06$  & $6.44\pm0.20$  & $0.18\pm0.02$  & $0.21\pm0.03$  & $1.906\pm0.017$  & $1.17$  \\
  27 & 62377 & $1.01\pm0.22$  & $0.53\pm0.05$  & $6.78\pm0.20$  & $0.15\pm0.01$  & $0.17\pm0.02$  & $1.837\pm0.011$  & $0.99$  \\
  28 & 64645 & $1.24\pm0.26$  & $0.57\pm0.05$  & $8.10\pm0.68$  & $0.09\pm0.01$  & $0.04\pm0.02$  & $1.165\pm0.013$  & $1.08$  \\
  29 & 66914 & $2.06\pm0.26$  & $0.75\pm0.05$  & $7.38\pm0.40$  & $0.10\pm0.01$  & $0.04\pm0.01$  & $0.825\pm0.007$  & $0.97$  \\
  31 & 71450 & $2.78\pm0.50$  & $0.77\pm0.10$  & $6.46\pm0.61$  & $0.11\pm0.02$  & $0.05\pm0.02$  & $0.760\pm0.019$  & $1.11$  \\
  32 & 73718 & $2.55\pm0.19$  & $0.80\pm0.04$  & $7.66\pm0.65$  & $0.10\pm0.01$  & $0.02\pm0.01$  & $0.762\pm0.007$  & $1.02$  \\
  34 & 78255 & $4.72\pm0.22$  & $0.84\pm0.11$  & $9.98\pm2.42$  & $0.12\pm0.01$  & $0.01\pm0.01$  & $0.893\pm0.032$  & $1.05$  \\
  35 & 80523 & $2.55\pm0.19$  & $0.60\pm0.03$  & $7.76\pm0.39$  & $0.14\pm0.01$  & $0.06\pm0.02$  & $1.547\pm0.009$  & $1.02$  \\
  37 & 85060 & $2.46\pm0.15$  & $0.67\pm0.03$  & $8.91\pm0.79$  & $0.13\pm0.01$  & $0.03\pm0.01$  & $1.296\pm0.009$  & $0.96$  \\
  38 & 87328 & $4.53\pm0.28$  & $0.63\pm0.05$  & $8.06\pm0.84$  & $0.13\pm0.01$  & $0.04\pm0.02$  & $1.202\pm0.017$  & $1.09$  \\
  39 & 89596 & $3.97\pm0.28$  & $0.61\pm0.05$  & $8.72\pm1.25$  & $0.10\pm0.01$  & $0.03\pm0.02$  & $1.064\pm0.018$  & $0.96$  \\
  40 & 91864 & $3.41\pm0.20$  & $0.50\pm0.03$  & $7.75\pm0.46$  & $0.10\pm0.01$  & $0.05\pm0.01$  & $1.171\pm0.008$  & $0.98$  \\
  42 & 96401 & $3.30\pm0.19$  & $0.51\pm0.05$  & $8.64\pm1.11$  & $0.08\pm0.01$  & $0.02\pm0.02$  & $1.073\pm0.015$  & $0.93$  \\
  43 & 98669 & $4.44\pm0.20$  & $0.61\pm0.08$  & $9.42\pm1.79$  & $0.09\pm0.01$  & $0.01\pm0.01$  & $0.954\pm0.020$  & $0.98$  \\
  45 & 103206 & $3.13\pm0.17$  & $0.39\pm0.03$  & $7.03\pm0.21$  & $0.14\pm0.01$  & $0.14\pm0.02$  & $1.998\pm0.008$  & $1.01$ \\
  46 & 105474 & $2.89\pm0.29$  & $0.39\pm0.06$  & $6.91\pm0.27$  & $0.13\pm0.01$  & $0.17\pm0.03$  & $1.884\pm0.016$  & $1.01$ \\
  \rowcolor{tablegray} 47 & 107742 & $7.42\pm0.32$  & $0.27\pm0.05$  & $7.28\pm0.37$  & $0.09\pm0.01$  & $0.11\pm0.03$  & $1.384\pm0.013$  & $1.10$  \\
  \rowcolor{tablegray} 48 & 110010 & $2.87\pm0.29$  & $-0.31\pm0.06$  & $7.14\pm0.39$  & $0.04\pm0.00$  & $0.20\pm0.09$  & $2.137\pm0.020$  & $1.10$  \\
  49 & 112279 & $5.94\pm0.42$  & $0.39\pm0.09$  & $6.42\pm0.18$  & $0.15\pm0.02$  & $0.38\pm0.05$  & $2.005\pm0.026$  & $1.11$  \\
  50 & 114547 & $5.97\pm0.24$  & $0.53\pm0.05$  & $6.38\pm0.10$  & $0.17\pm0.01$  & $0.30\pm0.02$  & $1.655\pm0.008$  & $1.00$  \\
  51 & 116815 & $3.95\pm0.24$  & $0.46\pm0.04$  & $7.86\pm0.60$  & $0.10\pm0.01$  & $0.05\pm0.02$  & $1.215\pm0.012$  & $1.04$  \\
  52 & 119083 & $3.59\pm0.23$  & $0.31\pm0.04$  & $7.84\pm0.64$  & $0.08\pm0.01$  & $0.05\pm0.02$  & $1.315\pm0.011$  & $1.15$  \\
  53 & 121352 & $3.75\pm0.20$  & $0.39\pm0.04$  & $6.98\pm0.23$  & $0.12\pm0.01$  & $0.13\pm0.02$  & $1.575\pm0.008$  & $0.92$  \\
  54 & 123620 & $3.45\pm0.25$  & $0.33\pm0.05$  & $6.77\pm0.26$  & $0.10\pm0.01$  & $0.15\pm0.03$  & $1.595\pm0.011$  & $1.01$  \\
  55 & 125888 & $6.06\pm0.24$  & $0.34\pm0.05$  & $6.57\pm0.14$  & $0.12\pm0.01$  & $0.24\pm0.03$  & $1.660\pm0.009$  & $1.03$  \\
  56 & 128156 & $6.17\pm0.33$  & $0.42\pm0.07$  & $6.38\pm0.15$  & $0.15\pm0.02$  & $0.32\pm0.04$  & $1.794\pm0.015$  & $1.06$  \\
  57 & 130425 & $4.22\pm0.26$  & $0.32\pm0.06$  & $6.43\pm0.12$  & $0.14\pm0.01$  & $0.41\pm0.04$  & $2.259\pm0.014$  & $1.12$  \\
  58 & 132693 & $5.90\pm0.23$  & $0.18\pm0.05$  & $6.21\pm0.08$  & $0.13\pm0.01$  & $0.61\pm0.04$  & $2.498\pm0.010$  & $1.03$  \\
  59 & 134961 & $7.67\pm0.38$  & $0.12\pm0.07$  & $7.03\pm0.23$  & $0.11\pm0.01$  & $0.31\pm0.06$  & $2.316\pm0.023$  & $1.09$  \\
  60 & 137229 & $5.82\pm0.27$  & $0.15\pm0.06$  & $6.43\pm0.12$  & $0.12\pm0.01$  & $0.51\pm0.05$  & $2.482\pm0.015$  & $1.09$  \\
  61 & 139498 & $7.32\pm0.33$  & $0.08\pm0.07$  & $6.45\pm0.14$  & $0.11\pm0.01$  & $0.52\pm0.07$  & $2.442\pm0.018$  & $1.04$  \\
  62 & 141766 & $5.67\pm0.25$  & $0.34\pm0.05$  & $6.34\pm0.09$  & $0.22\pm0.02$  & $0.77\pm0.05$  & $3.268\pm0.017$  & $1.10$  \\
  63 & 144034 & $2.60\pm0.21$  & $0.30\pm0.05$  & $6.15\pm0.08$  & $0.17\pm0.02$  & $0.68\pm0.05$  & $3.007\pm0.014$  & $1.02$  \\

  \hline
\end{tabular}
}
\end{center}
\begin{flushleft} {\scriptsize All errors represent 1$\sigma$
    uncertainties. a) photon index of the negative power law (PL). b)
    normalization of the negative power law in units of
    $\mathrm{photons~cm^{-2}s^{-1}}$ at 1 keV. c) normalization of the
    positive power law in units of $\mathrm{10^{-3}\
      photons~keV^{-1}cm^{-2}s^{-1}}$ at 1 keV. The data set referred in the text via the ID numbers are marked with gray color.}
\end{flushleft}
\end{minipage}
\end{table*}

\section{Discussions}\label{sec:discussion}

X-ray absorption and reprocessing provide us with crucial information on circumstellar medium in an HMXB.
Since the surroundings of the neutron star in the Vela X-1 system is an ideal site to investigate fueling processes to the accreting system as well as feedback processes of the accretion power, the circumstellar medium has been one of the main subject for observations of Vela X-1 \citep[e.g.][]{Nagase:1986, Sako:1999, Watanabe:2006}.
Based on the time variability of the absorption in Vela X-1, the relation between the structure of the stellar wind and the variability of the accreting object is discussed in this section.
In addition, we briefly discuss the radiation from the central engine based on the wide-band X-ray spectral data.

\subsection{Time Variability of Wind-fed Accretion System}

Typically, X-ray flux levels of wind-fed HMXBs show high variability, which is also seen in the light curves of Vela X-1 obtained with {\it Suzaku} (see Figs.~\ref{fig:xis_lc_284s}, \ref{fig:hxd_lc_284s}).
Similar flaring behavior in hard X-rays was already observed with {\it INTEGRAL} up to an intensity of $\sim 5\ \mathrm{Crab}$ at 20--40 keV \citep{Kreykenbohm:2008}.
Since hard X-rays also display similar variability to that of the soft band, the variability should be caused by changes of the intrinsic radiation from the neutron star, not due to an increase of the attenuation by the surrounding material around
the neutron star.

Inhomogeneity of the stellar wind is a possible origin of the time variability (See Section~1).
The luminosity $L_X$ by the wind-fed accretion varies as
\begin{equation}
L_X\propto \rho_\mathrm{wind}\ v_\mathrm{rel}{}^{-3}
\end{equation}
where $\rho_\mathrm{wind}$ and $v_\mathrm{rel}$ are the density and the relative velocity of the wind near the compact object \citep{Bondi:1944}.

It is probable that the wind of an OB star has highly clumpy structure in which density fluctuates with a factor of 3--10 originated from the line-driven instability of stellar UV photons \citep{Dessart:2003, Dessart:2005}.
The porosity of the wind is also inferred from symmetrical shapes of X-ray emission lines from isolated O-type stars obtained by high-resolution spectroscopy \citep{Oskinova:2006, Owocki:2006}.
Stellar winds in HMXBs are more complex due to the presence of the compact object and the strong X-ray radiation.
{\it ASCA} data of Vela X-1 in its eclipse phase suggested that the stellar wind of Vela X-1 consists of a number of dense clumps embedded in a tenuous, highly ionized plasma \citep{Sako:1999}.
\citet{Watanabe:2006} did not find any strong evidence for the clumpy structure of the wind in {\it Chandra} high-resolution data of Vela X-1, but confirmed that the wind velocity is suppressed near the neutron star due to the photoionization by the powerful X-rays, suggesting a strong impact of the X-rays on the dynamical structure of the wind.

\begin{figure}[htbp]
\begin{center}
\includegraphics[width=8.5cm]{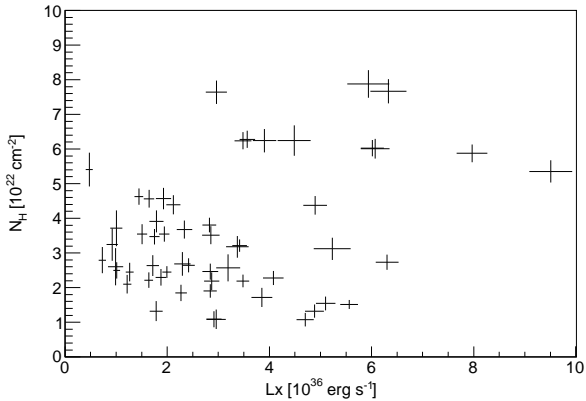}
\caption{Relation between the equivalent hydrogen column density $N_\mathrm{H}$ and the X-ray luminosity of Vela X-1. The two quantities are calculated based on the XIS/HXD simultaneous analysis described in \S\ref{subsec:wide_band_time_resolved_spectrum}. The luminosity is integrated energy over 0.1--100 keV, but the data fitting is based on 2.5--60 keV.}
\label{fig:plot_corr_nH_Lx}
\end{center}
\end{figure}

We examine here the {\it Suzaku} data based on the wind-clumping hypothesis.
We would expect some absorption features in soft X-ray bands associated with increase of the mass accretion rate if a dense clump dropped into the neutron star.
However, there is no significant trend of increase of X-ray absorption accompanied by luminous phases (Fig.~\ref{fig:plot_fit_xis_pl}).
In Figure~\ref{fig:plot_corr_nH_Lx}, we also show relation between $N_\mathrm{H}$ and the calculated X-ray luminosity $L_X$ both of which are based on the XIS/HXD simultaneous analysis.
There is no apparent correlation between the two quantities, and moreover $N_\mathrm{H}$ can be a very small value of $(1-2)\times 10^{22}\ \mathrm{cm^{-2}}$ even at high-luminosity states with $L_X \sim 5\times 10^{36}\ \mathrm{erg\ s^{-1}}$.

If the short-time variability of the flux and circumstellar absorption is a reflection of the wind clumpy structure, it is possible to estimate the physical properties of the individual clumps.
Indeed, the light curves obtained with {\it Suzaku} (see Fig.~\ref{fig:xis_lc_284s}) suggest the flare duration $t_f$ of 1--10 ks.
Assuming the relative velocity of the wind $v_\mathrm{rel} =400\ \mathrm{km\ s^{-1}}$ at the neutron star, we obtain the clump radius of $R_c \simeq v_\mathrm{rel}t_f/2 =(2-20)\times 10^{10}$ cm.
Since this radius is smaller than the accretion radius of Vela X-1 $R_\mathrm{acc}\sim 3\times 10^{11}$ cm, the neutron star can accrete the whole clump.
If $t_f=1$ ks and the flare luminosity $L_X=10^{37}\ \mathrm{erg\ s^{-1}}$ are assumed, the total energy $L_X t_f$ relates to the clump mass $M_c$, as
\begin{equation}
L_X t_f = \frac{GM_*M_c}{R_*} \quad\therefore M_c = \frac{L_X t_f R_*}{GM_*} = 4\times 10^{19}\ \mathrm{g},
\end{equation}
where $M_*$ and $R_*$ are the mass and the radius of the compact star, respectively.
Thus, we obtain the mean number density of the clump as
\begin{equation}
n_c = \frac{M_c}{m_\mathrm{p}}\frac{3}{4\pi R_c{}^3} = 7 \times 10^{11}\ \mathrm{cm^{-3}}
\end{equation}
and the radial column density $N_c$ of the clump $1.4\times 10^{22}\ \mathrm{cm^{-2}}$.
Consequently, the mean number density and the column density can be written as
\begin{gather}
\begin{split}
n_c =& 7\times 10^{11}\ \left( \dfrac{L_X}{10^{37}\ \mathrm{erg\ s^{-1}}} \right)\\
&\times \left(\dfrac{t_f}{1\ \mathrm{ks}} \right)^{-2} \left(\dfrac{v_\mathrm{rel}}{400\ \mathrm{km\ s^{-1}}} \right)^{-3}\ \mathrm{cm^{-3}}, \label{eq:clump_density_estimate}
\end{split} \\
\begin{split}
N_c =& n_cR_c =  1.4\times 10^{22}\ \left( \dfrac{L_X}{10^{37}\ \mathrm{erg\ s^{-1}}} \right)\\
&\times \left(\dfrac{t_f}{1\ \mathrm{ks}} \right)^{-1} \left(\dfrac{v_\mathrm{rel}}{400\ \mathrm{km\ s^{-1}}} \right)^{-2}\ \mathrm{cm^{-2}}.
\label{eq:clump_column_density_estimate}
\end{split}
\end{gather}

Since the number density of the wind around the neutron star which is estimated by the standard wind model \citep*{Castor:1975} is $n=3\times 10^9 \ \mathrm{cm^{-3}}$, the clump density estimated by equation~(\ref{eq:clump_density_estimate}) for a short flare of $t_f\sim 1$ ks is higher by two orders of magnitude.
However, hydrodynamical simulations by \citet{Blondin:1990, Blondin:1991} show existence of density fluctuation near the compact object with a factor of $\sim$100, and therefore such a dense clump may cause a short flare.
The column density of the clump is enough to detect by observation when the flare is bright and short.
The luminous flare at $t=110$ ks (see \S\ref{subsec:xis_lc}) is a candidate for such a flare.
We roughly estimate the column density of the clump that would generate the flare.
Adopting the luminosity $L_X=10^{37}\ \mathrm{erg\ s^{-1}}$ and the flare duration $t_f=400$ s estimated by visual inspection of the light curve (Fig.~\ref{fig:xis_lc_flare110ks}), we obtain the number density $n_c = 4\times 10^{12}\ \mathrm{cm^{-3}}$ and the column density $N_c=3.5\times 10^{22}\ \mathrm{cm^{-2}}$.
The difference of observed column densities just before the flare (ID: 47) and during the flare (ID: 48) is $\Delta N_\mathrm{H}=4\times 10^{22}\ \mathrm{cm^{-2}}$.
The good agreement of the decrease of the column density with the estimated value of the clump supports that the clump responsible for the absorption causes the following flare.

A larger wind clump can generate a longer flare.
We observed several moderate flares which have luminosities of $L_X\sim 5\times 10^{36}\ \mathrm{erg\ s^{-1}}$ and durations of $t_f\sim 5$ ks.
Such a flare corresponds to the number density $n_c = 1.4\times 10^{10}\ \mathrm{cm^{-3}}$ and the column density $N_c=1.4\times 10^{21}\ \mathrm{cm^{-2}}$.
Since this density is five times as dense as the value of the smooth wind model, the density enhancement is naturally explained by the line-driven instability of the stellar wind. 
The column density of the clump, however, is too small to detect by X-ray observations because of contamination of other absorption effects due to complicated wind structure close to the neutron star, as discussed in the next subsection.

In summary, the {\it Suzaku} observation of Vela X-1 shows consistency with the wind clumping responsible for the time variability of the X-ray radiation.
If the short flares of Vela X-1 are triggered by such clumps, they are likely to be very dense clumps around the neutron star, not the clumps which possibly exist in wide regions of the stellar wind.
However, other mechanisms such as the instability of the temporary accretion disk \citep{Taam:1989} or the so-called ``flip-flop instability'' \citep{Matsuda:1991} can possibly also explain the flaring behavior of the wind-fed accreting pulsar \citep[See also discussion in][]{Kreykenbohm:2008}.

\subsection{Enhanced Wind}

It has been well known that Vela X-1 shows strong variability of absorption with the orbital phase \citep{Pan:1994, Nagase:1986}.
{\it Chandra} observations also showed very different column densities of $N_\mathrm{H}=1.45\times 10^{22}\ \mathrm{cm^{-2}}$ at phase 0.25 and $N_\mathrm{H}=18.5\times 10^{22}\ \mathrm{cm^{-2}}$ at phase 0.5 \citep{Watanabe:2006}.
These absorption features are attributed to enhanced winds or accretion wakes around the neutron star, which are generated by competing effects including gravitational, rotational, and radiation pressure forces, and X-ray heating \citep{Blondin:1990, Blondin:1991}.
{\it Suzaku} allows us to trace the structure of such dense clouds close to the neutron star with finer time and spatial resolutions than those of the past observations.

The {\it Suzaku} observation covered the orbital phase between 0.175 and 0.31.
This early orbital phase has been thought to be a period in which the X-ray absorption is small, as observed with the {\it Chandra} at phase 0.25.
However, the light curves and the time-resolved spectral analysis of the {\it Suzaku} data shows relatively heavy absorption in the second half of the observation.
Similar absorption increase were seen by the {\it Tenma} satellite at the phase $\phi=0.25$ \citep[See Fig. 2 of][]{Nagase:1986}.
The {\it Suzaku} results also show strong fluctuation of absorption after $t=100$ ks (i.e. at orbital phase 0.27).

As a possible interpretation, we attribute this absorption feature to a bow shock generated by the interaction of the moving compact object with the supersonic stellar wind.
The hydrodynamic simulations by \citet{Blondin:1991} show such a structure emerging around the compact object.
According to the simulation, dense regions at the downstream of the shock will be observed as the heavy absorption at orbital phase $\phi\sim0.4$.
The related hump of absorption is in fact very similar to the absorption increase observed by {\it Suzaku} \citep[See Figures~2 and 8 from][]{Blondin:1991}. Although the orbital phase of the predicted and observed features are different, this interpretation still remains feasible, since the position of the bow shock greatly depends on the physical conditions of the binary system, including the binary separation, the wind properties, and so on.

\subsection{Spectral Features of the Central Engine}

\begin{figure}[htbp]
\begin{center}
\includegraphics[width=8.5cm]{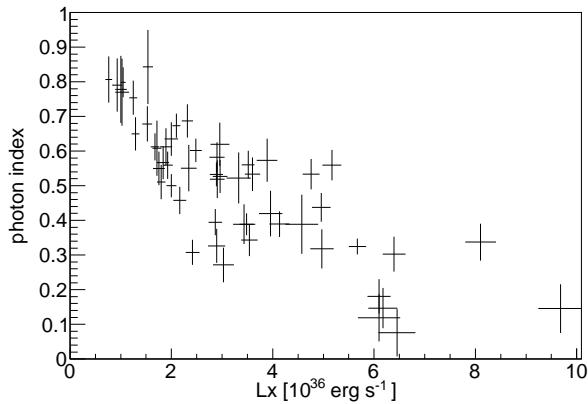}
\caption{Relation between the photon index $\Gamma$ and the X-ray luminosity. Two significantly hard and soft points are out of the $y$-range.}
\label{fig:relation_gamma_Lx}
\end{center}
\end{figure}

As shown in Figure~\ref{fig:plot_corr_gamma_flux}, we have found a correlation between the X-ray luminosity and the spectral hardness.
Instead of the X-ray flux, Figure~\ref{fig:relation_gamma_Lx} shows the same relation plotted with the luminosity in 2--10 keV of the source.
The source spectrum becomes harder as the luminosity increases.
On the assumption of the thermal Comptonization as the dominant process of the X-ray radiation, this correlation could have a natural interpretation: an enhanced accretion rate should lead both to an increase of the X-ray luminosity and to a more efficient Comptonization occurring in the accreted plasma.

The spectral features of Vela X-1---the power law with a quasi-exponential cutoff and the hardening with the luminosity---agree with the Comptonization hypothesis.
In order to confirm it, it is necessary to obtain more quantitative physical relations based on a more detailed emission model.
Such a model should incorporate a dynamical structure of the accretion flow and a magnetic field with physically reasonable values.
In our subsequently published paper \citep{Odaka:2013b}, detailed modeling using Monte Carlo simulations and implications of the observation based on the modeling are presented.

\section{Conclusions}
\label{sec:conclusions}

We studied the brightest wind-fed accreting neutron star Vela X-1 with the {\it Suzaku} X-ray observatory.
Due to the wide-band coverage and the low-noise performance of {\it Suzaku}, we have successfully obtained the time variability of the wide-band X-ray spectrum over a short timescale of 2 ks for the observation duration of 145 ks.
To exploit the high performance of {\it Suzaku} in the soft X-ray energy band, we have performed the data analysis in two steps: first we have analyzed the
data obtained with XIS only; and after that we conducted a broad-band analysis, which accounted for both XIS and HXD data.
The soft X-ray analysis has shown that:
\begin{itemize}
\item The ``low states'', in which the source luminosity becomes an order of magnitude lower than the averaged value, still shows X-ray pulsations, indicating that the accretion flow reaches the magnetic poles of the neutron star. In this state, the source displays a significant spectral softening with a photon index of $\sim$2.
\item The circumstellar absorption seems not to correlate with the X-ray luminosity except for a very hard flare which occurred at $t=110\ \mathrm{ks}$. This is consistent with the clumpy wind scenario to explain the time variability. However, other mechanisms such as the magnetospheric barrier effects cannot be excluded at this stage.
\item The equivalent hydrogen column density $N_\mathrm{H}$ increased in the second half of the {\it Suzaku} observation, which corresponds to the orbital phases of 0.25--0.3. This feature can be interpreted as a dense region where a bow shock is generated by interactions of the compact object with the stellar wind.
\end{itemize}

The broad-band X-ray analysis allowed us to obtain the following conclusions:
\begin{itemize}
\item The time-averaged wide-band spectrum  is well represented by the NPEX function, which is a combination of two different-slope power laws with a common exponential cutoff, with an obvious CRSF at 50 keV. However, we could not find any significant evidence for a CRSF at 25~keV in this spin-phase-averaged spectrum.
\item The time-resolved wide-band spectra between 2.5 keV and 50 keV are well described by the NPEX function. The time variability of the spectral properties should reflect variability of the physical conditions of the accreted plasma.
\item Data obtained from Vela X-1 show correlation of the photon index $\Gamma$ with the X-ray luminosity. On the assumption of the thermal Comptonization as the dominant process of the X-ray radiation, this correlation agrees with a natural physical relation that the optical thickness of the accreted plasma increases with the mass accretion rate onto the neutron star.
\end{itemize}

\acknowledgments

H.~Odaka and Y.~Tanaka have been supported by research fellowships of the Japan Society for the Promotion of Science for Young Scientists.
This work was supported in part by Global COE Program (Global Center of Excellence for Physical Sciences Frontier), MEXT, Japan.





\end{document}